\documentclass[aps,twocolumn,pra,tightenlines,floatfix,superscriptaddress]{revtex4-1}

\usepackage[breaklinks=true]{hyperref}
\usepackage{graphicx}
\usepackage[english]{babel}
\usepackage{amsmath}
\usepackage{amssymb}
\usepackage{times}

\begin{document}

\newcommand{\D}{\mathrm{D}}
\newcommand{\p}{\partial}
\newcommand{\Tr}{\mathrm{Tr}}
\renewcommand{\d}{\mathrm{d}}
\newcommand{\Ek}{E_\mathbf{k}}
\newcommand{\xik}{\xi_\mathbf{k}}
\newcommand{\sumk}{\sum_\mathbf{k}}

\title{Superfluidity and pairing phenomena in ultracold atomic Fermi
  gases in one-dimensional optical lattices, Part II: Effects of population imbalance}

\author{Jibiao Wang}
\affiliation{Guangdong Provincial Key Laboratory of Quantum Metrology and Sensing \& School of Physics and Astronomy, Sun Yat-Sen University (Zhuhai Campus), Zhuhai, Guangdong 519082, China}
\affiliation{Department of Physics and Zhejiang Institute of Modern Physics, Zhejiang University, Hangzhou, Zhejiang 310027, China}

\author{Lin Sun}
\affiliation{Department of Physics and Zhejiang Institute of Modern Physics, Zhejiang University, Hangzhou, Zhejiang 310027, China}

\author{Qiang Zhang}
\affiliation{Department of Physics and Zhejiang Institute of Modern Physics, Zhejiang University, Hangzhou, Zhejiang 310027, China}

\author{Leifeng Zhang}
\affiliation{Department of Physics and Zhejiang Institute of Modern Physics, Zhejiang University, Hangzhou, Zhejiang 310027, China}

\author{Yi Yu }
\affiliation{College of Chemical Engineering, Zhejiang University of Technology, Hangzhou, Zhejiang 310014, China}

\author{Chaohong Lee}
\affiliation{Guangdong Provincial Key Laboratory of Quantum Metrology and Sensing \& School of Physics and Astronomy, Sun Yat-Sen University (Zhuhai Campus), Zhuhai, Guangdong 519082, China}
\affiliation{State Key Laboratory of Optoelectronic Materials and Technologies, Sun Yat-Sen University (Guangzhou Campus), Guangzhou, Guangdong 510275, China}

\author{Qijin Chen}
\email[Corresponding author: ]{qchen@zju.edu.cn}
 \affiliation{Shanghai Branch, National Laboratory for Physical Sciences at Microscale and Department of Modern Physics, University
  of Science and Technology of China, Shanghai 201315, China}
\affiliation{Department of Physics and Zhejiang Institute of Modern Physics, Zhejiang University, Hangzhou, Zhejiang 310027, China}
\affiliation{Synergetic Innovation Center of Quantum Information and Quantum Physics, Hefei, Anhui 230026, China}

\date{\today}

\begin{abstract}

  In this paper, we study the effect of population imbalance and its
  interplay with pairing strength and lattice effect in atomic Fermi
  gases in a one-dimensional optical lattice. We compute various phase
  diagrams as the system undergoes BCS-BEC crossover, using the same
  pairing fluctuation theory as in Part I. We find widespread
  pseudogap phenomena beyond the BCS regime and intermediate
  temperature superfluid states for relatively low population
  imbalances. The Fermi surface topology plays an important role in
  the behavior of $T_\text{c}$. For large $d$ and/or small $t$, which yield
  an open Fermi surface, superfluidity can be readily destroyed by a
  small amount of population imbalance $p$. The superfluid phase,
  especially in the BEC regime, can exist only for a highly restricted
  volume of the parameter space. Due to the continuum-lattice mixing,
  population imbalance gives rise to a new mechanism for pair hopping,
  as assisted by excessive majority fermions, which may lead to
  significant enhancement of $T_\text{c}$ on the BEC side of the Feshbach
  resonance, and also render $T_\text{c}$ approaching a constant asymptote in
  the BEC limit, when it exists. Furthermore, we find that not all
  minority fermions will be paired up in BEC limit, unlike the 3D
  continuum case.  These predictions can be tested in future
  experiments.

\end{abstract}


\maketitle

\section{Introduction}

With multiple experimentally tunable parameters, ultracold atomic
Fermi gases and optical lattices have attracted enormous attention
\cite{Review,Bloch_RMP,Stringari_RMP}. Fermions in optical lattices
are often described by a Hubbard model
\cite{Bloch_RMP,Stringari_RMP,Micnas14AP}.  Among them, the
one-dimensional (1D) optical lattices have been realized
experimentally for a long time
\cite{Dyke11PRL,Martin12PRL,Feld11Nature}.  However, a proper
treatment of fermions in 1D optical lattices is not yet available,
since most theoretical in this regard addresses pure lattice cases
\cite{Gu07PRB,FeiguinPRB76,Rizzi,TormaPRL101,Roscilde12EPL,Buchleitner2012}. Theoretical studies on such a true 1D optical lattice in the
experimental sense have been scarce. Devreese \textit{et al.} studied
possible Fulde-Ferrell-Larkin-Ovchinnikov (FFLO) states \cite{FF,LO}
in such a 1D optical lattice
\cite{DevreesePRA83,Devreese2011,Devreese12MPLB}, but mostly
restricted to the BCS and crossover regimes.
Indeed, the superfluid and pairing physics in a 1D optical lattice has
not been adequately studied thus far.  In Part I of the present work
\cite{PartI}, we have systematically studied the behavior of BCS--BEC
crossover of atomic Fermi gases in a 1D optical lattice in the absence
of a population (and mass) imbalance. In particular, we have found
widespread pseudogap phenomena, which bear strong signatures in
single particle excitation spectrum and the superfluid density.

In this paper, we continue from Part I \cite{PartI} and study the
effects of population imbalance and its interplay with lattice
constant $d$ and lattice hopping parameter $t$, besides the
interaction strength and temperature, within the framework of the same
pairing fluctuation theory.  We find that the exponential behaviors of
the fermionic chemical potential $\mu$ and the pairing gap $\Delta$ as
a function of pairing strength in the BEC regime remain the same as in
the balanced case. The behavior of the superfluid transition
temperature $T_\text{c}$ is largely governed by the Fermi surface
topology. For large $d$ and/or small $t$, which lead to an open Fermi
surface, a small amount of population imbalance $p$ may readily
destroy superfluidity. Furthermore, the mixing between continuum and
discrete lattice dimensions has more profound consequences than in the
balanced case; the excessive majority fermions can assist pair
hopping, providing a new pair hopping mechanism, which dominates the
hopping via virtual pair unbinding \cite{NSR} in the BEC
regime. Together with the quasi-two dimensionality, which yields a
constant ratio $\Delta^2/\mu$ in the BEC limit, this new mechanism
leads to a constant asymptote for $T_\text{c}$ for a BEC superfluid (when a
BEC solution exists) in the presence of population imbalance.  We
shall present detailed $T$ -- $p$ (temperature versus polarization) phase
diagrams as the system undergoes the BCS-BEC crossover with different
lattice constants and hopping integrals, and focus on the finite
temperature and population imbalance effects, especially the pseudogap
phenomena \cite{Chen14Review,Mueller17RPP}. We shall also present
$T_\text{c}$ versus interaction strength $1/k_\text{F}a$ with varying lattice
constants $d$, population imbalances $p$ and hopping integrals $t$. As
these phase diagrams reveal, (i) the superfluid phase exists only in a
very restricted volume of the multi-dimensional parameter space; (ii)
the pseudogap phenomena widely exist; (iii) intermediate temperature
superfluidity is also a widespread phenomenon in the presence of
population imbalance, similar to the homogeneous case \cite{Chien06},
irrespective of the lattice constraint; (iv) a small population
imbalance may greatly enhance the superfluidity by raising $T_\text{c}$ on
the BEC side of the Feshbach resonance; (v) a BEC superfluid exists
only for a limited small volume in the parameter space of $(t,d,p)$,
and (vi) Not all minority fermions will be paired when a BEC
superfluid does exist.

\section{Theoretical Formalism}

In this section, we present briefly the theory, adapted for the
population imbalanced case, with spin dependent chemical potential
$\mu_\sigma$ and Green's functions $G_{0\sigma}(K)$ and $G_\sigma(K)$,
with the (pseudo)spins $\sigma=\uparrow,\downarrow$. We keep the same
notations as in Part I \cite{PartI}.

\subsection{Pairing fluctuation theory with a population imbalance}

The single particle dispersion is given by
$\xi_{\textbf{k}\sigma}=\textbf{k}^{2}_{\parallel}/2m+2t[1-\cos(k_{z}d)]-\mu_{\sigma}\equiv
\epsilon_{\textbf{k}}-\mu_{\sigma}$.
The bare Green's function is given by
$G^{-1}_{0\sigma}(K)=i\omega_{n}-\xi_{\textbf{k},\sigma}$, with the
self-energy $\Sigma_{\sigma}(K)=\sum_{Q}t(Q)G_{0\bar{\sigma}}(Q-K)$,
where $\bar{\sigma}=-\sigma$.  The $T$-matrix
$t(Q)=t_\text{sc}(Q)+t_\text{pg}(Q)$, where
$t_\text{sc}(Q)=-(\Delta_\text{sc}^{2}/T)\delta(Q)$ vanishes for $T>T_\text{c}$, and
$t_\text{pg}(Q)=U/[1+U\chi(Q)]$, with the pair susceptibility
$\chi(Q)=\sum_{K,\sigma}G_{0\sigma}(Q-K)G_{\bar{\sigma}}(K)/2$.  The
self-energy is given by
$\Sigma_{\sigma}(K)=\Sigma_{sc,\sigma}(K)+\Sigma_{pg,\sigma}(K)$,
where $\Sigma_{sc,\sigma}(K)= -\Delta_\text{sc}^{2}G_{0\bar{\sigma}}(-K)$,
and
$\Sigma_{pg,\sigma}(K)=\sum_{Q}t_\text{pg}(Q)G_{0\bar{\sigma}}(Q-K)$. At
$T\le T_\text{c}$, the BEC condition remains
$t^{-1}_\text{pg}(Q=0)=U^{-1}+\chi(0)=0$, and
$\Sigma_{pg,\sigma}(K)\approx -\Delta_\text{pg}^{2}G_{0\bar{\sigma}}(-K)$,
with $\Delta_\text{pg}^{2}\equiv-\sum_{Q}t_\text{pg}(Q)$.  Then the total
self-energy
$\Sigma_{\sigma}(K)\approx -\Delta^{2}G_{0\bar{\sigma}}(-K)$, where
$\Delta^{2}=\Delta_\text{sc}^{2}+\Delta_\text{pg}^{2}$. Finally, the full
Green's function becomes more complex due to population imbalance,
\begin{equation}
 G_{\sigma}(K)=\frac{u_{\textbf{k}}^{2}}{i\omega_{n}-E_{\textbf{k}\sigma}}+\frac{v_{\textbf{k}}^{2}}{i\omega_{n}+E_{\textbf{k}\bar{\sigma}}}\,,
\end{equation}
where $u_{\textbf{k}}^{2}=(1+\xi_{\textbf{k}}/E_{\textbf{k}})/2$, $v_{\textbf{k}}^{2}=(1-\xi_{\textbf{k}}/E_{\textbf{k}})/2$, $E_{\textbf{k}\uparrow}=E_{\textbf{k}}-h$, $E_{\textbf{k}\downarrow}=E_{\textbf{k}}+h$, and $E_{\textbf{k}}=\sqrt{\xi_{\textbf{k}}^{2}+\Delta^{2}}$, $\xi_{\textbf{k}}=\epsilon_{\textbf{k}}-\mu$, $\mu=(\mu_{\uparrow}+\mu_{\downarrow})/2$, $h=(\mu_{\uparrow}-\mu_{\downarrow})/2$. From the number constraint $n_{\sigma}=\sum_{K}G_{\sigma}(K)$, we can get the total fermion number density $n=n_{\uparrow}+n_{\downarrow}$ and the density difference $\delta n= n_{\uparrow}-n_{\downarrow}\equiv p n$,
\begin{eqnarray}
 n&=&\sum_{\textbf{k}}\Big[\Big(1-\frac{\xi_{\textbf{k}}}{E_{\textbf{k}}}\Big)+2\bar{f}(E_{\textbf{k}})\frac{\xi_{\textbf{k}}}{E_{\textbf{k}}}\Big]\,,\label{eq:LOFF_neqa}\\
 p n&=&\sum_{\textbf{k}}\Big[f(E_{\textbf{k}\uparrow})-f(E_{\textbf{k}\downarrow})\Big]\,,\label{eq:LOFF_neqb}
\end{eqnarray}
where $\bar{f}(x)=[f(x+h)+f(x-h)]/2$.
Similar to the $p=0$ case,  the extended gap
equation is given by
\begin{equation}
  \frac{m}{4\pi a}=\sum_{\textbf{k}}\Big[\frac{1}{2\epsilon_{\textbf{k}}}-\frac{1-2\bar{f}(E_{\textbf{k}})}{2E_{\textbf{k}}}\Big]+a_{0}\mu_\text{p}\,,\label{eq:gap}
\end{equation}
with $\mu_\text{p}=0$ at $T\leq T_{c}$.

The inverse $T$-matrix expansion \cite{Review} remains formally the
same as in the $p=0$ case, and all the coefficients are determined
automatically in the expansion process. Their concrete expressions are
given by Eqs.~(A4), (A5) and (A7) in the Appendix of Part I with the
Fermi distribution functions $f(x)$ and $f'(x)$ replaced by
$\bar{f}(x)$ and $\bar{f}'(x)$, respectively. The pseudogap equation
is the same,
\begin{equation}
 a_0\Delta_\text{pg}^{2}=\sum_{\textbf{q}}\frac{b(\tilde{\Omega}_{\textbf{q}})}{\sqrt{1+4\dfrac{a_{1}}{a_{0}}(\Omega_{\textbf{q}}-\mu_\text{p})}}\,,\label{eq:PG}
\end{equation}
with the pair dispersion
\[\tilde{\Omega}_{\textbf{q}}=\frac{\sqrt{a_{0}^{2}+4a_{1}a_0(\Omega_{\textbf{q}}-\mu_\text{p})}-a_{0}}{2a_{1}}.\]
The
pair density is given by $n_\text{p} = a_0\Delta^2$.

Equations (\ref{eq:LOFF_neqa})-(\ref{eq:PG}) form a closed set of
self-consistent equations, which can be used to solve for
($\mu_{\uparrow}$, $\mu_{\downarrow}$, $T^*$) with $\Delta=0$, for
($\mu_{\uparrow}$, $\mu_{\downarrow}$, $\Delta_\text{pg}$, $T_\text{c}$) with
$\Delta_\text{sc}=0$, and for ($\mu_{\uparrow}$, $\mu_{\downarrow}$,
$\Delta$, $\Delta_\text{pg}$) at $T<T_\text{c}$. Here the pair formation
temperature $T^*$ is approximated by the mean-field $T_\text{c}$, and the
order parameter $\Delta_\text{sc}$ is derived from
$\Delta_\text{sc}^2=\Delta^2-\Delta_\text{pg}^2$. 

\subsection{Stability analysis}

In the presence of population imbalance, not all solutions of
Eqs.~(\ref{eq:LOFF_neqa})-(\ref{eq:PG}) are stable. The stability
analysis can be done following Ref.~\cite{Stability}, as we summarize
here. Consider the thermodynamic potential $\Omega_\text{S}$, which
consists of the fermionic ($\Omega_\text{F}$) and bosonic ($\Omega_\text{B}$)
contributions,
\begin{eqnarray}
 \Omega_\text{S}&=&\Omega_\text{F}+\Omega_\text{B}\,,\\
 \Omega_\text{F}&=&-\frac{\Delta^{2}}{U\,\,}+\sum_{\textbf{k}}(\xi_{\textbf{k}} -E_{\textbf{k}})-T\sum_{\textbf{k},\sigma}\ln\,(1 + e^{-E_{\textbf{k}\sigma}/T})\,,\nonumber\\
 \Omega_\text{B}&=&a_{0}\mu_\text{p}\Delta^{2}_\text{pg}+T\sum_{\textbf{q}}\ln(1-e^{-\tilde{\Omega}_{\textbf{q}}/T})\,.\nonumber
\end{eqnarray}
The stability condition of population imbalanced Sarma phase
\cite{Sarma63} against phase separation (PS) can be simply expressed
as
\begin{equation}
 \frac{\partial^{2}\Omega_\text{S}}{\partial\Delta^{2}}=2\sum_{\textbf{k}}\frac{\Delta^2}{E_{\textbf{k}}^2}\Big[\frac{1-2\bar{f}(E_{\textbf{k}})}{2E_{\textbf{k}}}+\bar{f}'(E_{\textbf{k}})\Big]>0\,,\label{eq:sta}
\end{equation}
where $\bar{f}'(x)=d\bar{f}(x)/dx$. This condition is equivalent to
the positive definiteness of the particle number susceptibility matrix
$\{\partial n_{\sigma}/\partial\mu_{\sigma'}\}$
\cite{PWY05,Stability}, which represents a form of generalized
compressibility.

\subsection{Superfluid density}

Similar to the $p=0$ case, the superfluid ``density'' $(n_\text{s}/m)$, can
also be derived using the linear response theory, following earlier
works \cite{ChenPRL98,Chien06,Stability}. 

For the present contact potential, the superfluid density is given by
\begin{equation}
  \left(\frac{n_\text{s}}{m} \right)_{i}     = 2  \sum_\mathbf{k} \frac{\Delta_\text{sc}^2}{\Ek^2} \left[ \frac{1-\bar{f}(\Ek)}{2\Ek} +\bar{f}^\prime(\Ek)\right]
                                        \left( \dfrac{\partial\xik}{\partial{k}_i}\right)^2 ,
  \label{eq:Ns}
\end{equation}
where $i=x,y,z$ and $\bar{f}'(x) = \text{d}\bar{f}(x)/\text{d}x$.

As we will see below, the behavior of the superfluid density can
becomes very usual for $p\ne 0$. Nevertheless, we expect the $T$
dependence of both $({n_\text{s}}/{m})_\parallel$ and $({n_\text{s}}/{m})_z$ are
close to each other.

\subsection{Asymptotic behavior in the deep BEC regime}

Unlike the $p=0$ case \cite{PartI}, in the presence of a population
imbalance $p\neq 0$, the BEC limit is more complicated, as one can no
longer obtain a complete analytical solution without resorting to
numerics. However, one can still greatly reduce the complexity of the
equations, as follows.

For $p = (n_\uparrow -n_\downarrow)/n$, we consider $p>0$, without
loss of generality. The excessive majority fermions require
$\mu_\uparrow >0 $ throughout the BCS--BEC crossover, whereas $\mu$ to
leading order is roughly given by its balanced counterpart in the BEC
limit, where the two-body physics dominates. Then $\mu_\downarrow$ is
given by $\mu_\downarrow = 2\mu -\mu_\uparrow$. The size of
$\mu_\uparrow >0 $ is determined by $p$, and
$\mu_\downarrow \approx 2\mu \rightarrow -\infty$, so that
$f(\Ek^\downarrow) = f(\xik^\downarrow) = 0$. The Fermi function
$f(\Ek^\uparrow)$ no longer vanishes exponentially, and will lead to
corrections to the equations above. Nevertheless, this Fermi function
places a small finite energy and momentum cutoff, so that we have
$\Ek\approx |\mu|$ to the leading order in many occasions. Thus to
leading order corrections,  the equation for total number
density now becomes
\begin{eqnarray}
  \label{eq:nBECp}
  (1-p)n &=& -\frac{m\Delta^2}{4\pi \mu d} -\frac{np\Delta^2}{2\mu^2}\\
  \Delta &=& \sqrt{\frac{4\pi|\mu|d(1-p)n}{m}} \left(1-\frac{\pi dnp}{\mu m} \right)\,.
  \label{eq:gapBECp}
\end{eqnarray}
Interestingly, the leading correction to $\Delta^2$ is independent of
$1/k_\text{F}a$, given by $8(\pi dn/m)^2(1-p)p$, which vanishes when
$p=0$. So is the correction term in Eq.~(\ref{eq:nBECp}).

Expanding $\Ek^\uparrow$, we have
\begin{equation}
  E_\mathbf{k}^\uparrow = \Ek-h \approx \xik^\uparrow -\frac{\Delta_\mathbf{k}^2}{2\mu} \approx \xik^\uparrow+ \frac{4\pi dn_\downarrow}{m} \,.
  \label{eq:E-xi}
\end{equation}
Note that the second term is again a constant for given $p$,
independent of $1/k_\text{F}a$, precisely because
$\Delta^2/\mu\rightarrow \text{const}$. For this reason, the
difference
$E_\mathbf{k}^\uparrow - \xik^\uparrow = 4\pi dn_\downarrow/m$ will
\emph{not} approach 0 in the BEC limit, unlike the case in 3D
continuum.

 The equation of number difference is given by
\begin{eqnarray}
  pn &=& \sumk f(\Ek^\uparrow)= \sumk f(\xik^\uparrow+\frac{4\pi dn_\downarrow}{m})   \equiv  \frac{mt}{\pi^2d}I_1\,.
  \label{eq:nupBEC}
\end{eqnarray}
Here the dimensionless integral $I_1$ depends on $\mu_\uparrow$ and
$T$.

In comparison with the $p=0$ case, the gap equation now also contains
an extra term which is of the same order as the leading term in the
BEC limit, namely,
\begin{equation}
  \sumk\frac{f(\Ek^\uparrow)}{2\Ek}  \approx \frac{1}{2|\mu|}\sumk f(\Ek^\uparrow) = \frac{pn}{2|\mu|}\,,
  \label{eq:chiCorrection}
\end{equation}
Thus without this term, the leading order chemical potential is given by
$
\mu_0 = -t e^{d/a},
$
the same as in the $p=0$ case, since the two-body physics dominates
the deep BEC regime. 
The gap equation can now be simplified in a fashion similar to the $p=0$
case, and we obtain
\begin{equation}
  \mu = \mu_0 + 2t + \frac{2\pi dn_\uparrow}{m} \,,
  \label{eq:muBECp}
\end{equation}
formally identical to the expression for $p=0$. Plugging
Eq.~(\ref{eq:muBECp}) into Eq.~(\ref{eq:gapBECp}), we can obtain the
gap $\Delta$. Note that for given $(t,d,p)$ in the deep BEC regime,
Eqs.~(\ref{eq:muBECp}) and (\ref{eq:gapBECp}) completely determines
$\mu$ and $\Delta$ as a function of $1/k_\text{F}a$.

As discussed in Part I, the exponential behavior of $\mu$ and $\Delta$
as a function of $1/k_\text{F}a$ is an important feature of the quasi-two
dimensionality of the present system; the ratio $\Delta^2/\mu $
approaches a constant, independent of pairing strength. As we shall
see below, this has important consequences. The (2nd and 3rd)
correction terms in Eq.~(\ref{eq:muBECp}) are also constants.

Finally, to solve for $T_\text{c}$ (and $\mu_\uparrow$), we need to simplify
the expressions for the dispersion of the pairs. Defining
$ \sumk f(\xik^\uparrow)
\equiv \dfrac{mt}{\pi^2d} I_2 $, then the coefficient $a_0$ is given
by
\begin{eqnarray}
 n_\text{p} &=& a_0 \Delta^2 = 
                    \frac{n}{2} - \frac{1}{2}\sumk f(\Ek^\uparrow) +\frac{1}{2}\sumk [f(\Ek^\uparrow) - f(\xik^\uparrow)]\nonumber\\
               &=& n_\downarrow - \frac{mt}{2\pi^2d}(I_2 - I_1) \,.
                   \label{eq:a0BECp}
\end{eqnarray}
Note here both the integrals $I_1$ and $I_2$ depend only on
$\mu_\uparrow$ and $T$, which are independent of the pairing strength
in the BEC limit. Both will vanish when $p=0$. However, in the
presence of population imbalance, $I_2-I_1$ will not vanish in the BEC
limit due to Eq.~(\ref{eq:E-xi}). Therefore, the pair density, $n_\text{p}$,
will approach a constant BEC asymptote, which is smaller than
$n_\downarrow$ for $p > 0$. Namely, \emph{not} all minority fermions
will be paired up.

The coefficient $a_1$ is now given by
\begin{equation}
a_1\Delta^2 = \frac{m^2t}{8\pi^3 d^2 n_\downarrow}(I_2-I_1) + \frac{1}{4|\mu|} \Big(n_\downarrow + \frac{m^2t^2}{\pi^3d^2n_\downarrow } I_3 \Big)\,,
\label{eq:a1BECp}
\end{equation}
where the integral
$I_3 =\dfrac{\pi^2d}{2mt^2}\sumk\epsilon_\mathbf{k}[f(\xik^\uparrow) -
f(\Ek^\uparrow)] $.  Again, for $p=0$, all the $I$'s vanish, so that
Eq.~(\ref{eq:a1BECp}) recovers the $p=0$ result,
$a_1\Delta^2 = n/8|\mu|$.  It is a dramatic difference that a finite
population imbalance contributes a finite, constant, first term on the
right hand side of Eq.~(\ref{eq:a1BECp}).

After some lengthy but straightforward derivation, we obtain
\begin{equation}
  B_\parallel = \frac{1}{4m} + \frac{1}{4n_\text{p}} \Big[\frac{t}{2\pi^2d}(3I_2+I_1) - \frac{mt^2}{2\pi^3d^2n_\downarrow}I_4 \Big]\,,
  \label{eq:BBECp}
\end{equation}
where $I_4=\dfrac{\pi^2d}{2m^2t^2}\sumk [f(\xik^\uparrow) -
f(\Ek^\uparrow)] k_\parallel^2 $. The first
term is the $p=0$ result, while the rest is the contribution of population
imbalance. Here we have kept only the leading order terms and dropped
terms of order $1/\mu$ or higher. The pair density $n_\text{p}$ is to be
replaced with  Eq.~(\ref{eq:a0BECp}).

The pair hopping integral $t_\text{B}$ is given by
\begin{eqnarray}
  t_\text{B}  &=& \frac{t^2}{n_\text{p}}\Bigg\{\frac{m}{2\pi^2d}\Big(I_5-I_6+I_7- \frac{mt}{\pi d n_\downarrow} I_8\Big) \nonumber\\
       &&{}+ \frac{n_\downarrow}{2|\mu|}\bigg(1 -\frac{8}{\pi}I_5 -\frac{4t^2m^2}{\pi^3d^2n_\downarrow^2}I_9 \bigg)\Bigg\}
          \label{eq:BzBECp}
\end{eqnarray}
where
\begin{eqnarray}
I_5 &=& \dfrac{\pi^2d}{mt}\sumk f(\Ek^\uparrow) \cos (k_zd),\nonumber\\
I_6 &=& \dfrac{\pi^2d}{mt} \sumk f(\xik^\uparrow) \cos (k_zd),\nonumber\\
I_7 &=& -\dfrac{4\pi^2d}{m}\sumk f'(\xik^\uparrow) \sin^2 (k_zd),\nonumber\\
I_8 &=& \dfrac{\pi^2d}{mt}\sumk [f(\xik^\uparrow) -
f(\Ek^\uparrow)]\sin^2 (k_zd) ,\nonumber\\
I_9 &=& \dfrac{\pi^2d}{2mt^2}\sumk \epsilon_\mathbf{k}
[f(\xik^\uparrow) - f(\Ek^\uparrow)]\sin^2 (k_zd).\nonumber
\end{eqnarray}

For $p=0$, all integral $I$'s vanish so that Eq.~(\ref{eq:BzBECp})
reduces to the $p=0$ result, $t_\text{B} = t^2/2|\mu|$. As in
Eq.~(\ref{eq:BBECp}), here $n_\text{p}$ is to be replaced with
Eq.~(\ref{eq:a0BECp}). Once again, population imbalance leads to the
first term in the brackets in Eq.~(\ref{eq:BzBECp}), which is a
constant of interaction strength and thus becomes the dominant
term. This will dramatically change the behavior of the $T_\text{c}$
solution.

Equation (\ref{eq:muBECp}) completely determines $\mu$, and then
Eq.~(\ref{eq:gapBECp}) is used to fully fix the gap $\Delta$, for
given $1/k_\text{F}a$ in the deep BEC regime.  Since the quantities $n_\text{p}$,
$a_1$, $B_\parallel$ and $t_\text{B}$ rely only on $\mu_\uparrow$ and $T$
(with corrections of order $O(1/\mu)$), then $\mu_\uparrow$ and $T_\text{c}$
can be obtained via solving the pseudogap equation (\ref{eq:PG}) along
with the number difference Eq.~(\ref{eq:nupBEC}), with
$\Delta_\text{pg}=\Delta$. Note that Eq.~(\ref{eq:PG}) depends only on the
product $a_0\Delta^2$ and the ratio $a_0/a_1$, but not on the value of
$\Delta$. The fact that the leading terms of $a_0\Delta^2$,
$a_1\Delta^2$, $a_0/a_1$, $B_\parallel$ and $t_\text{B}$ are all independent
of $\mu$ or $\Delta$ in the presence of a population imbalance implies
that $\mu_\uparrow$ and $T_\text{c}$, along with these quantities, all
approach their respective interaction-independent BEC asymptotes,
which depend only on $(t,d,p)$.

\section{Numerical Results and Discussions}

\begin{figure}
  \centerline{
    \includegraphics[clip,width=3.20in]{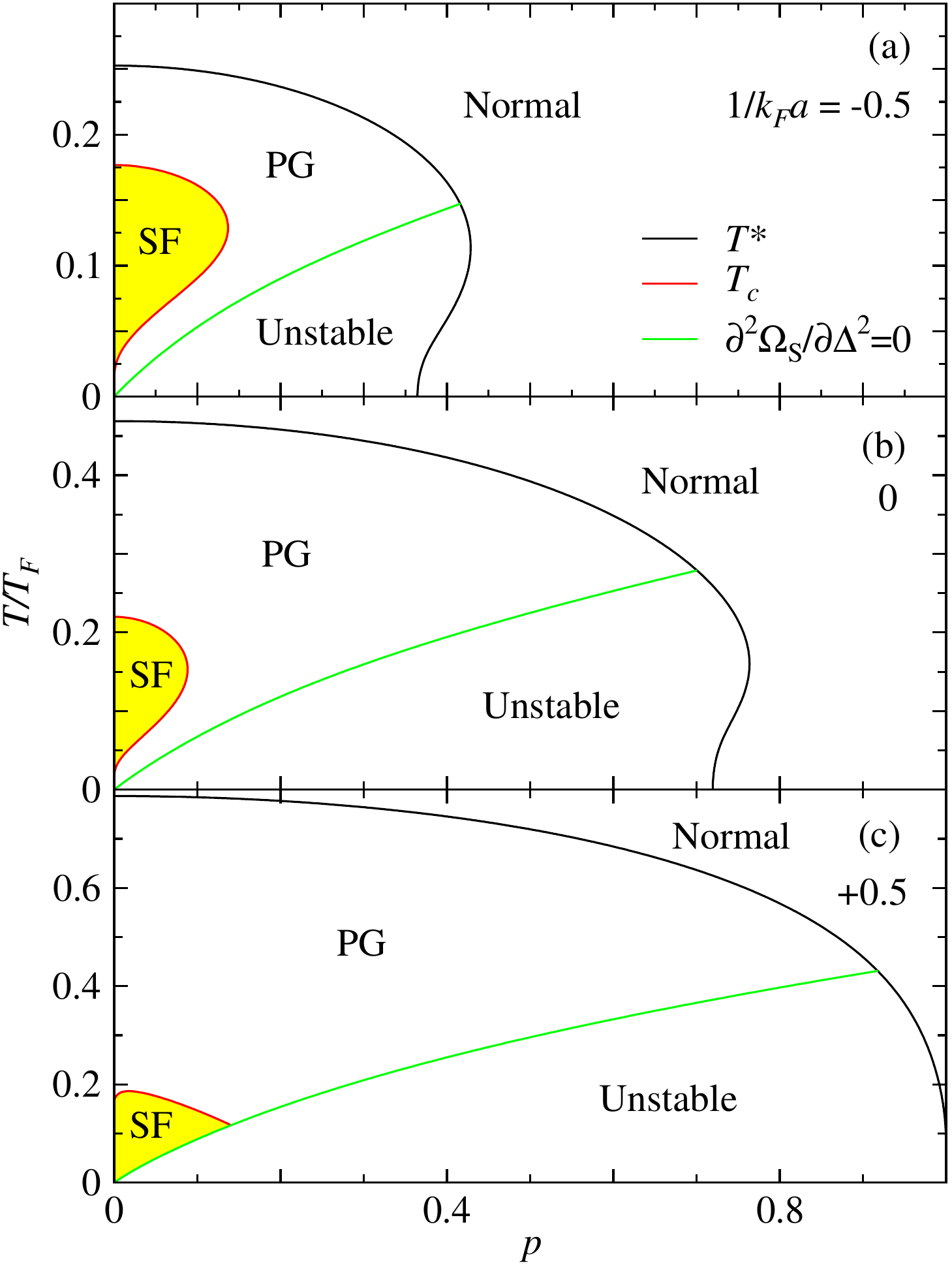}}
\caption{Evolution of the phase diagram in the $T$ -- $p$ plane with
  $(t/E_\text{F},k_\text{F}d) = (1,1)$, for (a) $1/k_\text{F}a= -0.5$, (b) 0, and (c) 0.5,
  corresponding to near-BCS, unitary, and near-BEC cases,
  respectively. Here ``PG" and ``SF" indicate the pseudogapped normal
  state and superfluid, respectively. The stability condition of
  Eq. (\ref{eq:sta}) is violated in the ``Unstable" region. }
\label{fig:t1kFd1}
\end{figure}

In this subsection, we present our results in the presence of a
population imbalance, while the parameters $(t,d,1/k_\text{F}a)$ vary.

For our numerical calculations, we define Fermi momentum
$k_\text{F}=(3\pi^{2}n)^{1/3}$ and Fermi energy
$E_\text{F}\equiv k_\text{B}T_\text{F}=\hbar^{2}k_\text{F}^{2}/2m$, as given by a
homogeneous, balanced, noninteracting Fermi gas with the same total
number density $n$ in 3D.

\subsection{Effect of population imbalance on BCS--BEC crossover}

\subsubsection{An unphysical  nearly isotropic case: $t/E_\mathrm{F}=1$ and $k_\mathrm{F}d=1$}

\begin{figure}
    \centerline{\includegraphics[clip,width=3.2in]{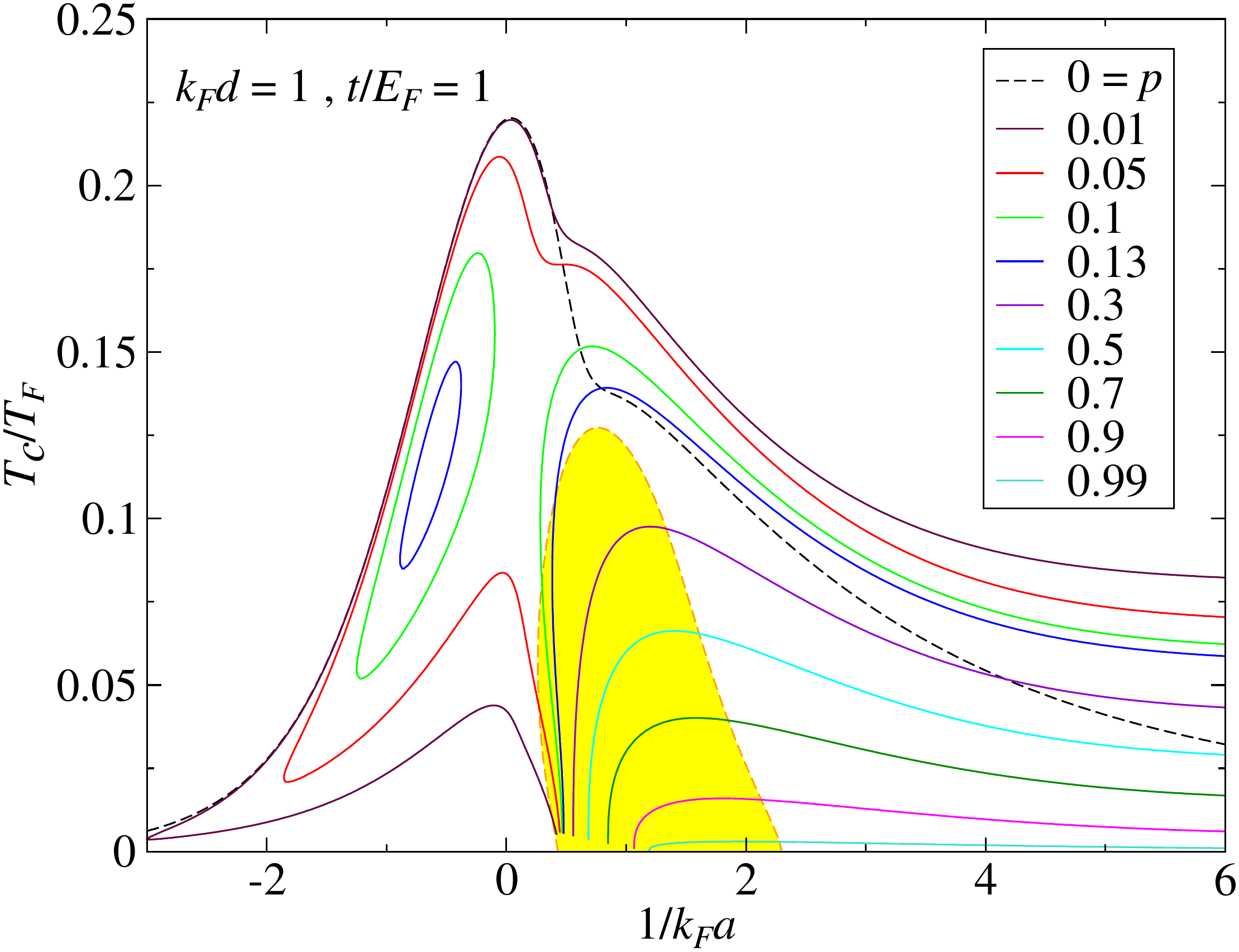}}
\caption{$T_\text{c}$ -- $1/k_\text{F}a$ phase diagram for different $p$ at fixed $k_\text{F}d=1$ and $t/E_\text{F}=1$. The $T_\text{c}$ solutions inside the shaded region are unstable.}
\label{fig:kFd1t1}
\end{figure}

First, we consider the case $t/E_\text{F}=1$ and $k_\text{F}d=1$, which is not
physically accessible, but provides a nearly spherical Fermi surface
in the noninteracting limit \cite{Zhang17SR}, and thus may serve to
make contact with the 3D homogeneous case \cite{Stability}.
Shown in Fig.~\ref{fig:t1kFd1} is the evolution of the phase diagram in
the $T$ -- $p$ plane for three representative pairing strengths in the
(a) near-BCS, (b) unitary, and (c) near-BEC regimes, respectively. The
phase diagram in each case consists of a small intermediate
temperature, Sarma (i.e., polarized) superfluid phase (yellow shaded,
labeled ``SF''), a large pseudogapped normal phase (``PG''), an
unpaired normal Fermi gas phase (``Normal''), as well as an unstable
phase (``Unstable''), which often gives way to phase separation
\cite{notesonFFLO}. Considering the different vertical scales, the
superfluid phase has roughly comparable phase space volumes for the
three cases, more or less similar to its homogeneous counterpart in 3D
free space, as shown in Figs. 6 and 7 in Ref.~\cite{Stability}.  Here
the (in)stability condition (green line) is given by
Eq.~(\ref{eq:sta}).  Indeed, For $k_\text{F}d=1$, we have $\pi/d \gg k_\text{F}$, so
that the confinement in $k_z$ has only a minor impact on the momentum
distribution. In addition, similar to the 3D homogeneous situation,
the unitary case has the highest $T_\text{c}$ at $p=0$ among all three cases,
and there exists no stable Sarma superfluid at $T=0$ when $p\neq 0$
for the cases considered ($1/k_\text{F}a \leq 0.5$). At $T=0$, the $p=0$ and
$p\neq 0$ cases are \emph{not} continuously connected in the BCS and
unitary regimes. A zero $T$ polarized superfluid solution exists only
in the deep BEC regime \cite{Chien06,Stability}. At the same time, the
(red) $T_\text{c}$ curve intersects with the (green) instability boundary for
the near-BEC case. And in the deep BEC regime, the instability line
intersects with the $p$ axis at a finite value, indicating the
existence of a stable zero $T$ polarized Sarma superfluid.

Now we turn to the effect of population imbalance on the behavior of
$T_\text{c}$ throughout the BCS-BEC crossover. Keeping $T_\text{c}$ as the function,
there are still four independent control variables, $p$, $1/k_\text{F}a$, $t$
and $d$, which can yield many different facets of the very rich phase
space. In this section, we shall only present a few very informative
phase diagrams.

Shown in Fig.~\ref{fig:kFd1t1} is the calculated $T_\text{c}$ -- $1/k_\text{F}a$
phase diagram for different $p$ from 0.01 to 0.99 at fixed $k_\text{F}d=1$
and $t/E_\text{F}=1$. For comparison, we also plot the $p=0$ curve (black
dashed). This figure bears a lot of similarity with that for the
simple 3D homogeneous case, shown in Ref.~\cite{Chien06}. For both
cases, there exist intermediate temperature superfluids from the BCS
to the near-BEC regime. This unusual phase has a higher and a lower
$T_\text{c}$ for a given $1/k_\text{F}a$. At the same time, for intermediate levels
of $p$ (0.1 and 0.13 shown here), the $T_\text{c}$ curve splits into two
branches, and the left branch shrinks to zero and disappears as $p$
further increases.  The $T_\text{c}$ solutions inside the yellow shaded
region do not satisfy the stability condition of Eq. (\ref{eq:sta}),
and hence are unstable. The difference comes mainly on the BEC side.
As $1/k_\text{F}a$ increases into the BEC regime, for our present case, $T_\text{c}$
decreases, which is qualitatively consistent with the $p=0$ cases
shown in Figs.~1 and 2 of Part I \cite{PartI}, reflecting the lattice
effect on pair hopping.

The most surprising feature in Fig.~\ref{fig:kFd1t1} is that $T_\text{c}$ for
$p=0$ decreases faster, and thus intersects with the $p\ne 0$
curves. This means we can get a higher $T_\text{c}$ by allowing a small
population imbalance on the BEC side of the Feshbach
resonance. Indeed, as we have shown analytically in
Eq.~(\ref{eq:BzBECp}), due to population imbalance, an additional
mechanism for pair hopping kicks in; a pair can hop to its neighboring
site via exchanging only the majority fermion component of a pair with
an excessive majority fermion that is already present on the
neighboring site, leaving the previous majority fermion component
behind. In this way, the minority fermion component glides through the
sites whereas the majority fermions do not necessarily have to hop.
Note here that a ``site'' in the lattice dimension corresponds
actually to a 2D plane, which guarantees that there are always
excessive majority fermions available on the neighboring ``site'',
when $p\ne 0$, in the thermodynamic limit. This is a consequence of
lattice-continuum dimensional mixing. The presence of a transverse
continuum dimension is crucial for this to happen. Due to this new
pair hopping mechanism, $t_\text{B}$ approaches a constant in the BEC limit,
and so does $T_\text{c}$. Indeed, as one can see, the $T_\text{c}$ curves already
flatten out towards BEC.

\subsubsection{Realistic cases with smaller $2mtd^2 < 1$}

\begin{figure}
\centerline{
  \includegraphics[clip,width=3.20in]{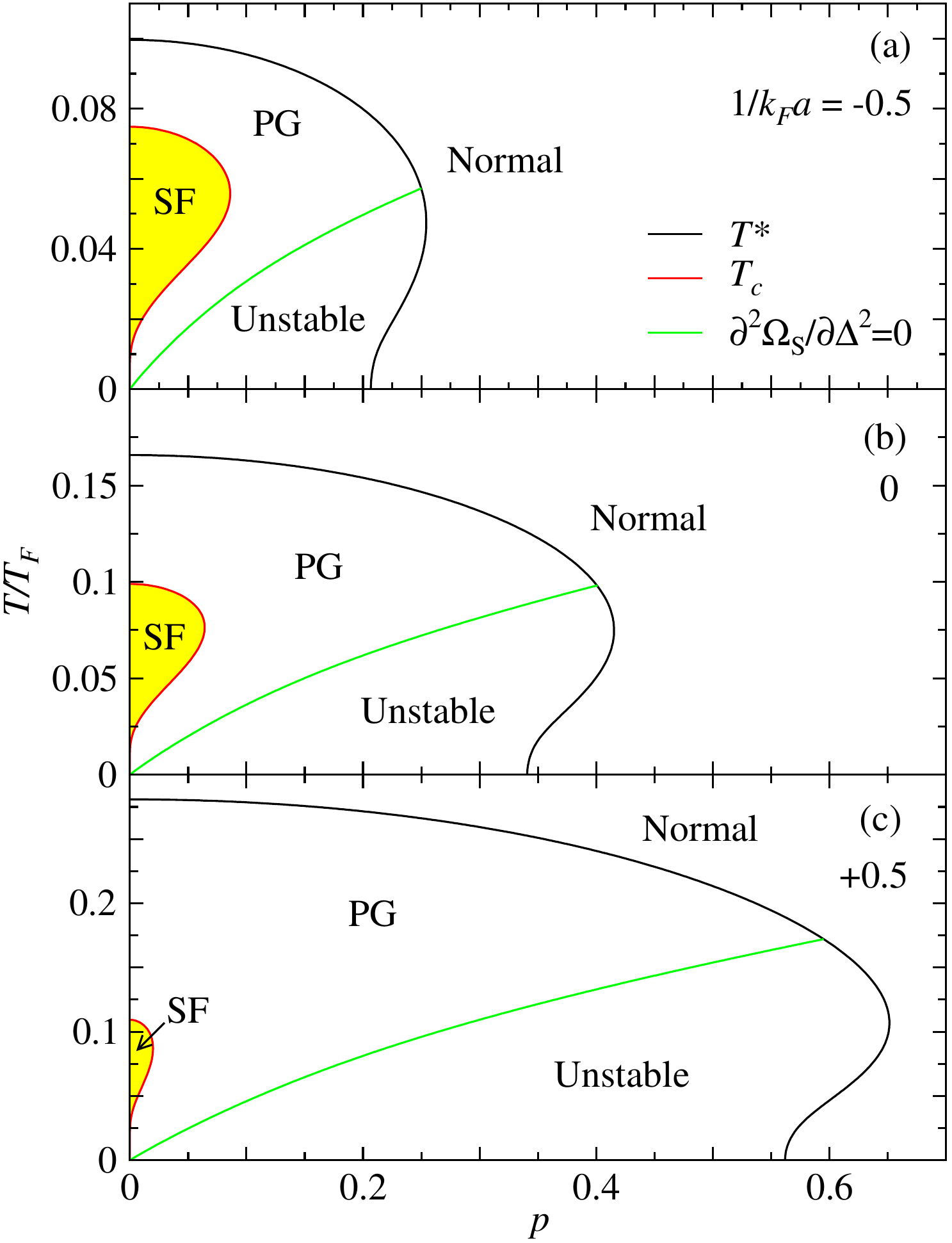}}
\caption{Evolution of the $T$ -- $p$ phase diagram with $t/E_\text{F}=0.05$ and
  $k_\text{F}d=2$ for different pairing strengths. Other parameters are the
  same as in Fig.~\ref{fig:t1kFd1}}
\label{fig:t0.05kFd2}
\end{figure}

Now we consider more realistic cases which are accessible
experimentally, as constrained by the condition $2mtd^2 < 1$. Shown in
Fig.~\ref{fig:t0.05kFd2} are the $T$ -- $p$ phase diagrams with
$(t/E_\text{F}, k_\text{F}d) = (0.05, 2)$, for the same values of $1/k_\text{F}a$ as in
Fig.~\ref{fig:t1kFd1}. In comparison, we observe that the reduced
$(t,d)$ or $td^2$ has led to significant reduction on $T_\text{c}$ and the
phase space volumes of the superfluid (``SF'') and paired (``PG'' and
``Unstable'') phases. This reduction reveals that the small $t$ and
relatively large $d$ are detrimental to both superfluidity and
pairing. The most dramatic effect is the rapid shrink of the SF phase
as $1/k_\text{F}a$ increases towards the BEC regime. Further more, the $T_\text{c}$
curve no longer intersects with the instability line. This suggests
that for finite $p>0$, there is \emph{no} superfluidity at $T=0$ even
in the deep BEC regime, for the present choice of $(t,d)$. On the
other hand, the superfluid solution for $p=0$ always exist
\cite{PartI}; in that case, the area of the SF phase does not
completely vanish even though it may become very small. Here one may
also notice that the unitary case no longer has the highest
$T_\text{c}$. This is because the maximum $T_\text{c}$ for $k_\text{F}d=2$ has shifted away
from unitarity towards the BEC side in the 1D optical lattice
\cite{PartI}. As one can expect, the smaller $t$ and larger $d$ make the
system quasi-2D, giving rise to stronger pairing fluctuations and thus
reduced $T_\text{c}$.

\begin{figure}
\centerline{\includegraphics[clip,width=3.4in]{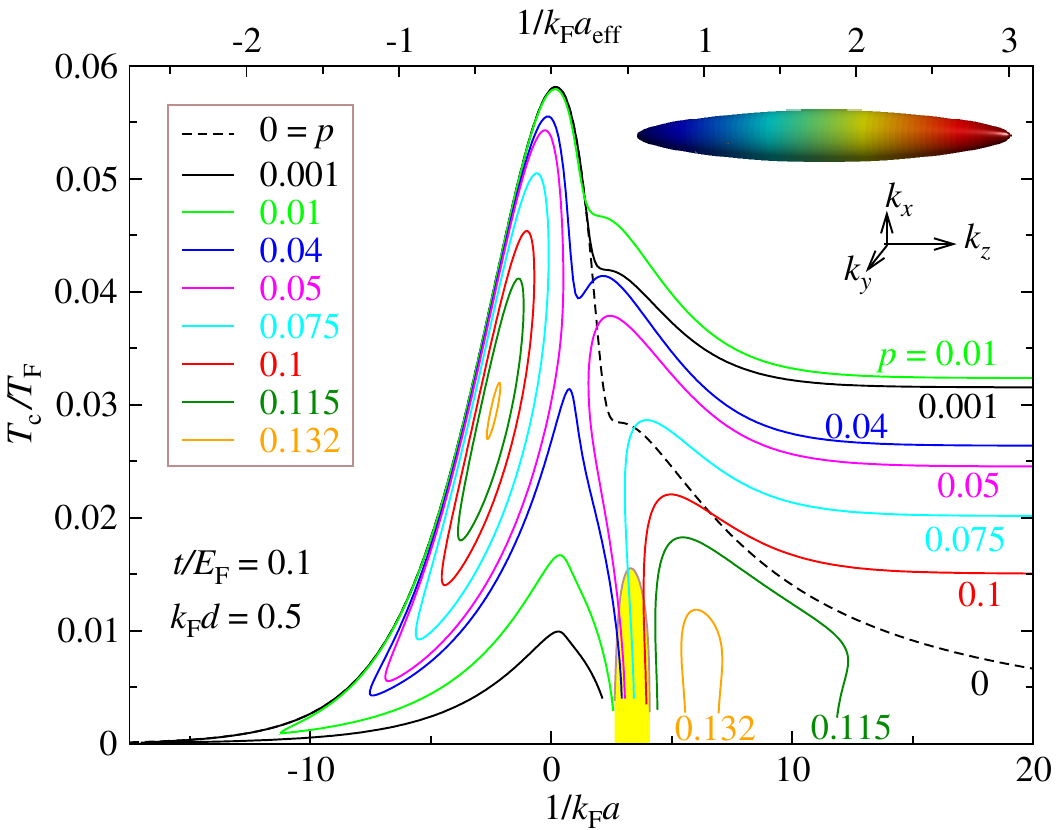}}
\caption{$T_\text{c}$ -- $1/k_\text{F}a$ phase diagram for different $p$ from 0 to
  0.132 (as labeled) at fixed $k_\text{F}d=0.5$ and $t/E_\text{F}=0.1$, showing
  dramatic effect of $t$ and $d$, when compared with
  Figs.~\ref{fig:kFd1t1} and \ref{fig:t0.1d1.5}. The $T_\text{c}$ solution
  within the yellow shaded area is unstable. Also labeled on the top
  axis is the effective parameter $1/k_\text{F}a_\text{eff}$. Shown in
  the inset is a 3D plot of the Fermi ellipsoid.}
\label{fig:kFd0.5t0.1}
\end{figure}

In analogy to Fig.~\ref{fig:t1kFd1}, we show in
Fig.~\ref{fig:kFd0.5t0.1} a realistic case with $t/E_\text{F} = 0.1$ and
$k_\text{F}d=0.5$. With this reduced $t$ and $d$, the Fermi surface is an
elongated ellipsoid in the noninteracting limit, as shown in the
inset.  Plotted here is $T_\text{c}$ as a function of $1/k_\text{F}a$ for different
$p$ from 0 to 0.132, as labeled next to the color coded curves. Also
labeled on the top axis is the effective parameter
$1/k_\text{F}a_\text{eff} = \sqrt{2mt}d/k_\text{F}a$, as defined in
Part I \cite{PartI} and Ref.~\cite{1DOLshort}. This parameter is
certainly closer to the $1/k_\text{F}a$ parameter of the 3D
homogeneous case \cite{Chien06}.  Similar to that in
Fig.~\ref{fig:t1kFd1}, the superfluid $T_\text{c}$ solution within the small
yellow shaded area is unstable.  In addition, the lower branch $T_\text{c}$
vanishes somewhere close to but on the BEC side of unitarity. In
comparison with Fig.~\ref{fig:t1kFd1}, however, the overall $T_\text{c}$ is
strongly suppressed by a factor of 4. This reduced $T_\text{c}$ is mainly
caused by the small $t$ and small $d$, which brings the noninteracting
chemical potential down dramatically to $\mu \approx 0.276E_\text{F}\approx
E_\text{F}/4$. The other main difference is that the population imbalance $p$
cannot go to a high value as it does in Fig.~\ref{fig:t1kFd1}, before
$T_\text{c}$ disappears completely. While the $T_\text{c}$ curve can still persists
into the BEC limit for $p \le 0.1$, it bends back for $p = 0.115$ and
forms a superfluid dome in the near-BEC regime.  The superfluid
phase quickly shrinks when $p$ increases further, and then disappears
for $p \gtrsim 0.132$.

To understand the difference between Figs. \ref{fig:kFd0.5t0.1} and
\ref{fig:t1kFd1}, we note that the elliptical Fermi surface in
Fig.~\ref{fig:kFd0.5t0.1} can be rescaled more or less into a sphere;
this allows for some similarities in the $T_\text{c}$ curves. However, as
pairing strength increases and the pairing gap becomes large, the pair
occupation number $v_\mathbf{k}^2$ (and hence the fermion momentum
distribution) will soon feel the confinement of the limited momentum
space in the lattice direction. As a consequence, the excessive
majority fermions will no longer be evenly distributed in all
directions (after the rescaling). This causes pairing more difficult
in the BEC regime and thus leads to a dome shape of the superfluid
phase. It also explains why $p$ cannot be large before superfluidity
disappears.

\begin{figure*}
\centerline{\includegraphics[clip,height=3.3in]{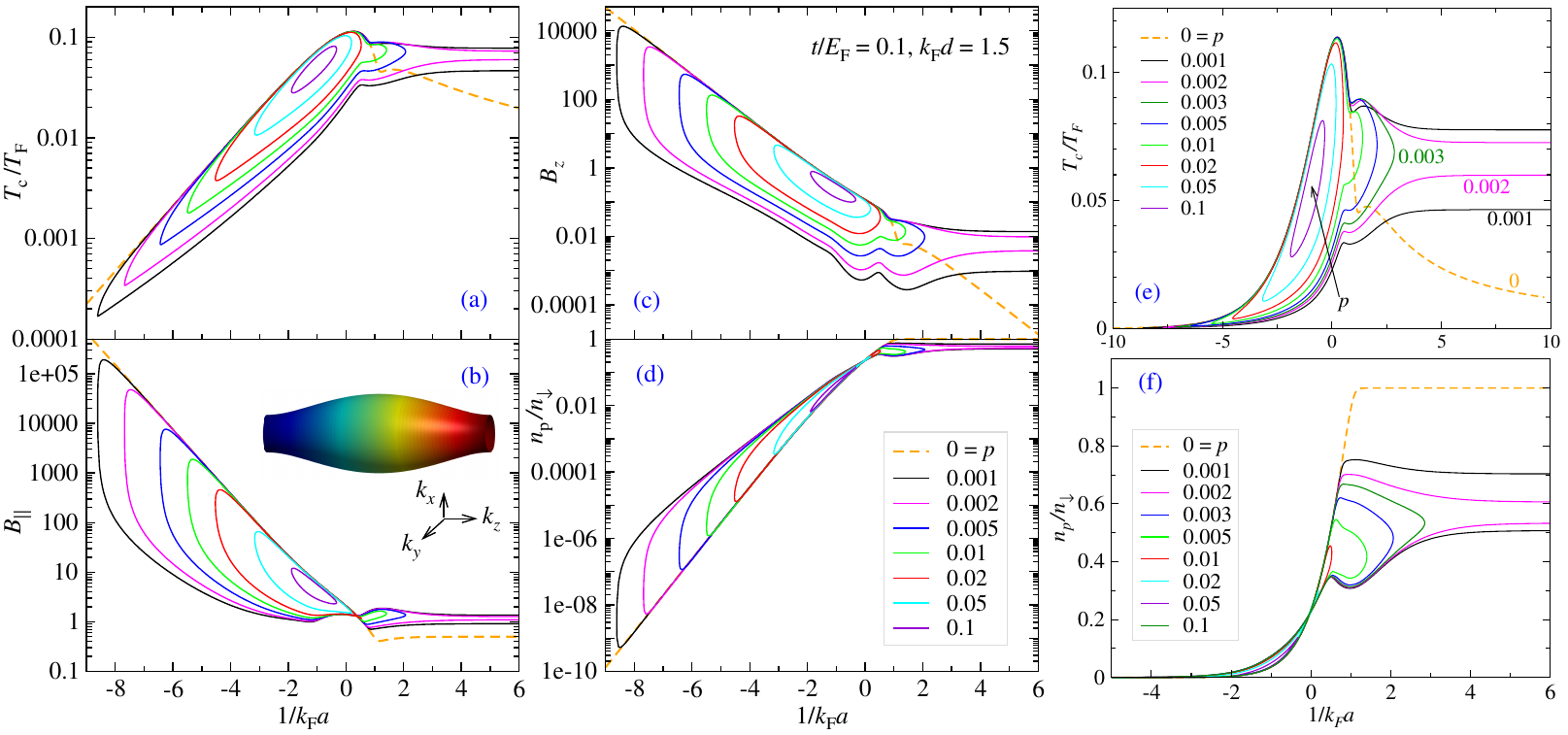}}
\caption{Behavior of (a, e) $T_\text{c}/T_\text{F}$, (b) $B_\parallel$, (c) $B_z$ (in
  units of $1/2m$) and (d, f) $n_\text{p}/n_\downarrow$ as a function of
  $1/k_\text{F}a$ for different $p$ from 0 to 0.1 at fixed $k_\text{F}d=1.5$ and
  $t/E_\text{F}=0.1$. The color coding for panels (a)-(d) are the same. }
\label{fig:t0.1d1.5}
\end{figure*}

Next, we keep $t/E_\text{F}=0.1$ but increase the lattice spacing $d$ to
$k_\text{F}d = 1.5$ so that the pairs feel more strongly the restriction of
$|k_z| \le \pi/d$.  Shown in Fig.~\ref{fig:t0.1d1.5} are the behaviors
of (a,e) $T_\text{c}$, the coefficients (b) $B_\parallel$ and (c) $B_z$, and
(d,f) the pair fraction $n_\text{p}/n_\downarrow$ (all at $T_\text{c}$) as a
function of $1/k_\text{F}a$ for a series of $p$ from 0 to 0.1. 
The Fermi surface now has open ends at $k_z =\pm \pi/d$, as shown in
the inset of panel (b). It can no longer become nearly spherical by
momentum rescaling. This inevitably shall lead to a bigger difference
from Fig.~\ref{fig:t1kFd1}. The curves in panels (a)-(d) are plotted
in a semi-log scale, making the exponential dependence of $T_\text{c}$ on
$1/k_\text{F}a$ for $p=0$ in the BCS regime self-evident as a straight line
(orange dashed). It turns out that the coefficients $B_\parallel$,
$B_z$ and pair density $n_\text{p}$ all bear similar exponential
dependencies. Panels (e) and (f) are plotted in linear scales. In the
presence of a finite imbalance $p$, as the interaction strength
\emph{decreases}, $T_\text{c}$ follows the $p=0$ curve until it hits the
lower threshold, at which it curves back into a lower branch of
$T_\text{c}$. Similar behaviors happen to $B_\parallel$, $B_z$ and $n_\text{p}$ as
well. On the other hand, on the BEC side of the Feshbach resonance,
$B_z$ approaches a constant for $p\ne 0$, (and $B_\parallel$ differs
substantially from its $p=0$ value). Accordingly, $T_\text{c}$ approaches a
constant BEC asymptote, and so does $n_\text{p}$.  All superfluid solutions
in Fig.~\ref{fig:t0.1d1.5} are stable.  Panels \ref{fig:t0.1d1.5}(d,f)
reveal that the pair density $n_\text{p}$ is higher along the lower branch of
$T_\text{c}$ than the upper branch, as expected.  We note that
$n_\text{p}/n_\downarrow < 1$, indicating that not all minority fermions form
pairs even in the deepest BEC limit, in contrast to the 3D continuum
case.
The BEC asymptotic behaviors are governed by
Eqs.~(\ref{eq:a0BECp})-(\ref{eq:BzBECp}).

Similar to Fig.~\ref{fig:kFd1t1}, in both Figs.~\ref{fig:kFd0.5t0.1} and
\ref{fig:t0.1d1.5} the $p=0$ curve for $T_\text{c}$ quickly drops with
increasing $1/k_\text{F}a$ and intersects with the $p\ne 0 $ curves. Namely,
in these physically accessible cases, our earlier finding about the
enhancement of $T_\text{c}$ by population imbalance remains valid.

In comparison with Fig.~\ref{fig:kFd1t1}, a big qualitative difference
is that there is no moderate level of $p$ in Fig.~\ref{fig:t0.1d1.5}
such that the $T_\text{c}$ curve splits into a left and a right branch. In
addition, due to the big difference between Fermi surfaces of these
two cases, the lower $T_\text{c}$ here does not vanish in the neighborhood of
unitarity, but rather either extends all the way to the BEC limit (for
small $p \leq 0.002$) or curls up and joins the upper $T_\text{c}$ before it
enters the deep BEC regime (for $p \geq 0.003$). The $T_\text{c}$ curve for
$p=0.003$ can extends into the BEC regime up to $1/k_\text{F}a = 2.854$ or
$1/k_\text{F}a_\text{eff} = 1.354$ . Furthermore, here we do not find
the counterpart $T_\text{c}$ curve that is similar to the $t/E_\text{F}=0.115$ case
in Figs.~\ref{fig:kFd0.5t0.1}. Therefore, while one may find a BEC
superfluid for large $p$ up to nearly unity in Fig.~\ref{fig:kFd1t1},
it is not possible for the quasi-2D case in
Fig.~\ref{fig:t0.1d1.5}. Indeed, the superfluid solution will
disappear from the entire phase space when $p > 0.124$ for the present
parameters $(t/E_\text{F},k_\text{F}d)=(0.1,1.5)$. In other words, superfluidity now
exists only in a small portion of the phase space; for small $t$ and
relatively not so small $d$, the superfluid phase can be easily
destroyed by a small amount of population imbalance. In addition, a
deep BEC superfluid exists only for very low $p$ as well.  Reducing
$t$ and/or increasing $d$ further may destroy completely the
superfluid phase even in the deepest BEC limit. Therefore, one needs
to reduce $d$ and/or increase $t$ to have a superfluid with a
relatively large $p$, as will be shown soon below.

\begin{figure}
\centerline{\includegraphics[clip,width=3.2in]{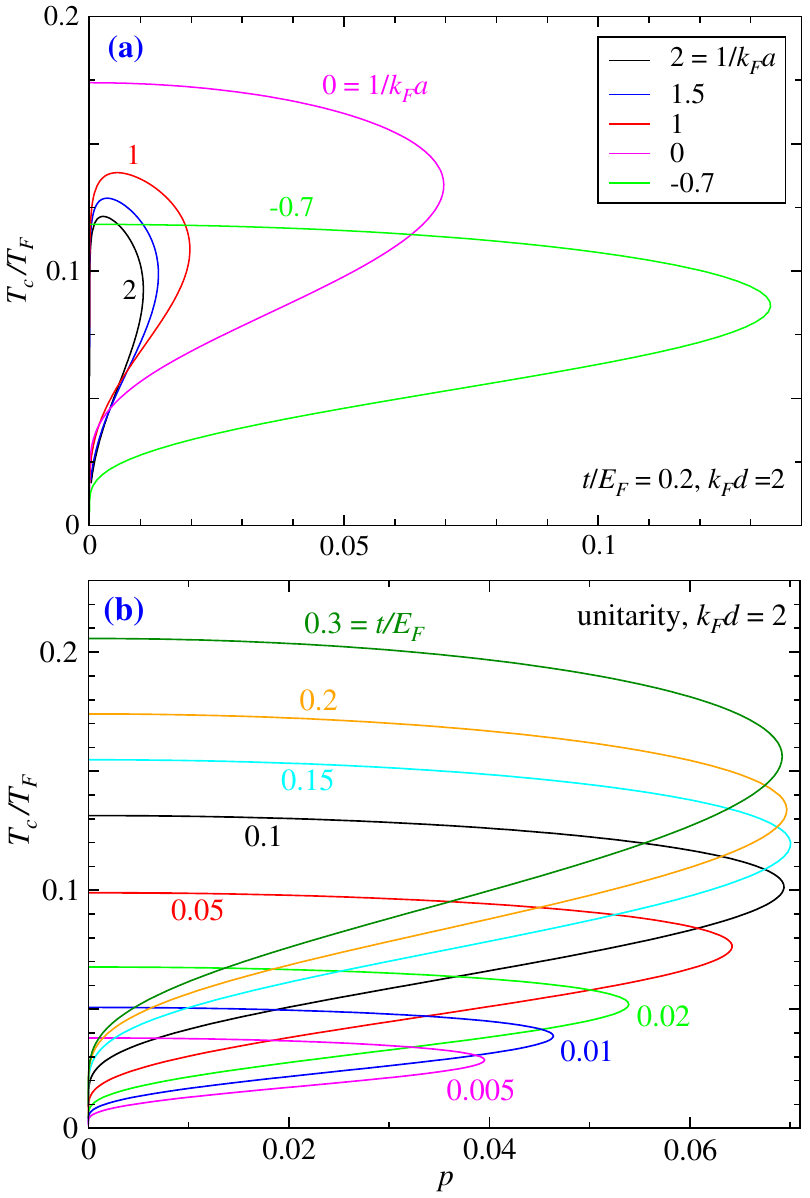}}
\caption{$T_\text{c}$ -- $p$ phase diagram with $k_\text{F}d=2$ (a) for different
  $1/k_\text{F}a$ from -0.7 to 2 at fixed $t/E_\text{F}=0.2$, and for (b) different
  values of $t/E_\text{F}$ from 0.005 to 0.3 (as labeled) at unitarity.}
\label{fig:kFd2t0.2-p-InvkFa}
\end{figure}


We notice that the enhancement of $T_\text{c}$ or superfluidity by population
imbalance occurs mainly on the BEC side of unitarity. To show this
more explicitly, we plot in Fig.~\ref{fig:kFd2t0.2-p-InvkFa}(a) the
behavior of $T_\text{c}$ as a function of $p$ at a series of pairing
strengths for fixed $(t/E_\text{F},k_\text{F}d)=(0.2,2)$. While one may find a
maximum allowable range of $p$ around $1/k_\text{F}a= -0.7$, and a maximum
$T_\text{c}$ at unitarity, these two cases do not see the enhancement effect,
since for both cases, $T_\text{c}$ reaches its maximum at $p=0$. In contrast,
for $1/k_\text{F}a = 1$, 1.5 and 2, as $p$ increases from 0, $T_\text{c}$
experiences an initial rapid jump from its $p=0$ value to a much
higher value at $p>0$, and then slowly drops down and bends back
towards $p=0$. There exists a significant range of $p$ in which $T_\text{c}$
is larger than its $p=0$ counterpart. The back-bending behavior of
$T_\text{c}$ versus $p$ is consistent with the intermediate temperature
superfluidity with an upper and lower $T_\text{c}$. The much reduced maximum
$p$ for these cases demonstrates that a superfluid solution exists
only for small $p$ on the BEC side of unitarity for the current
$(t,d)$ combination.

\subsection{Influence of $t$ and $d$ on the superfluid phase diagrams}

\subsubsection{$T$ -- $1/k_\text{F}a$ phase diagrams for different $t$ and $d$}

The effect of increasing $t/E_\text{F}$ on this phase diagram is shown in
Fig.~\ref{fig:kFd2t0.2-p-InvkFa}(b), where $T_\text{c}$ vs $p$ at unitarity
is plotted for a series of $t$ at $k_\text{F}d=2$. The maximum $T_\text{c}$ at $p=0$
increases with $t$, but the maximum reachable $p$ seems to saturate
for $t/E_\text{F}>0.1$. 

The evolution of superfluid phase from Fig.~\ref{fig:kFd1t1}, to
Fig.~\ref{fig:t0.1d1.5} and Fig.~\ref{fig:kFd2t0.2-p-InvkFa} tells
that in the presence of a population imbalance, the superfluid phase
volume decreases quickly and then disappears completely as the system
evolves into the quasi-2D regime.

\begin{figure}
\centerline{\includegraphics[clip,width=3.2in]{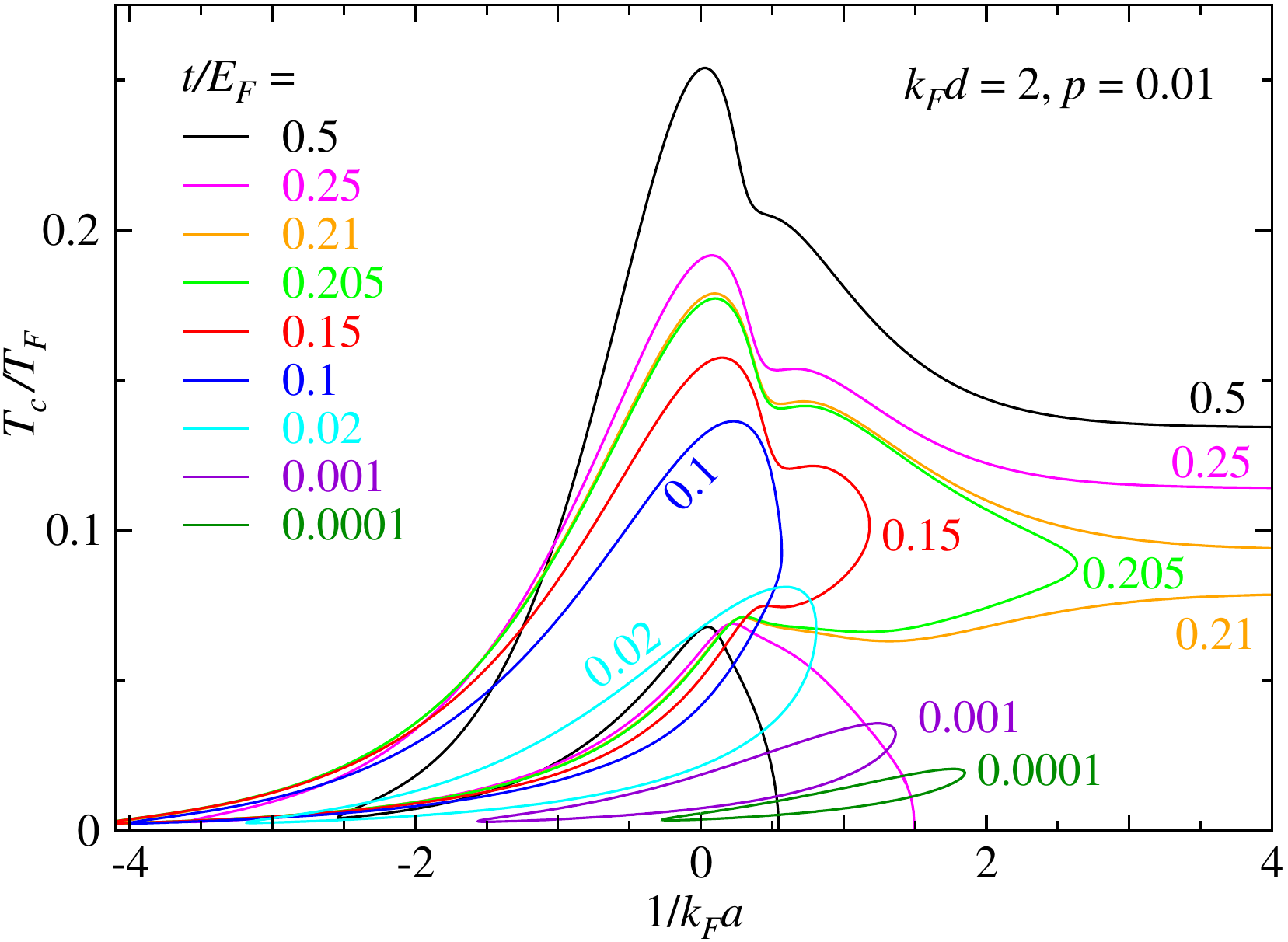}}
\caption{Behavior of $T_\text{c}$ as a function of $1/k_\text{F}a$ at fixed $k_\text{F}d=2$
  and $p=0.01$, but for different values of $t/E_\text{F}$, as labeled. It
  becomes more 3D-like as $t$ increases.}
\label{fig:Tc-InvkFa-t}
\end{figure}

If we allow ourselves to use somewhat larger range of $t$, we will
obtain the $T_\text{c}$ curves shown in Fig.~\ref{fig:Tc-InvkFa-t} as a
function of $1/k_\text{F}a$. Here we fix $p=0.01$ and $k_\text{F}d=2$, but vary
$t/E_\text{F}$ from 0.0001 up to 0.5, as labeled. For small $t/E_\text{F} \leq 0.1$,
we have a simple closed loop. Both $T_\text{c}$ and the size of the loop
increases as $t$ grows. For $t/E_\text{F}=0.15$ (red) and 0.205 (green
curve), the $T_\text{c}$ loop extends into the BEC regime, but still cannot
reach the deep BEC limit; the $T_\text{c}$ curve turns back somewhere on the
BEC side of unitarity, and form a closed cycle.  As $t$ increases
further, for $t/E_\text{F} \geq 0.21$, the $T_\text{c}$ curves successfully extend
all the way into the BEC limit. For $t/E_\text{F} = 0.21$ (orange curve),
both the upper and lower $T_\text{c}$ branches extend to $1/k_\text{F}a =
+\infty$. However, for $t/E_\text{F} \geq 0.25$ (black and pink curves), the
lower $T_\text{c}$ branch bends downwards around unitarity and vanishes at an
intermediate pairing strength, somewhere on the BEC side of
unitarity. In such a case, there exists a stable homogeneous polarized
superfluid in the BEC regime at $T=0$, similar to the case for a
simple 3D continuum. For $k_\text{F}d=2$, our calculation reveals that the
Fermi surface has two open ends at $k_z = \pm \pi/d$ for
$t/E_\text{F}\le 0.21$, whereas it becomes a closed ellipsoid again for the
large $t/E_\text{F} \ge 0.25$ cases. The corresponding $T_\text{c}$ behavior for the
latter cases is similar to that found in Fig.~\ref{fig:kFd1t1}.

\begin{figure}
\centerline{\includegraphics[clip,width=3.2in]{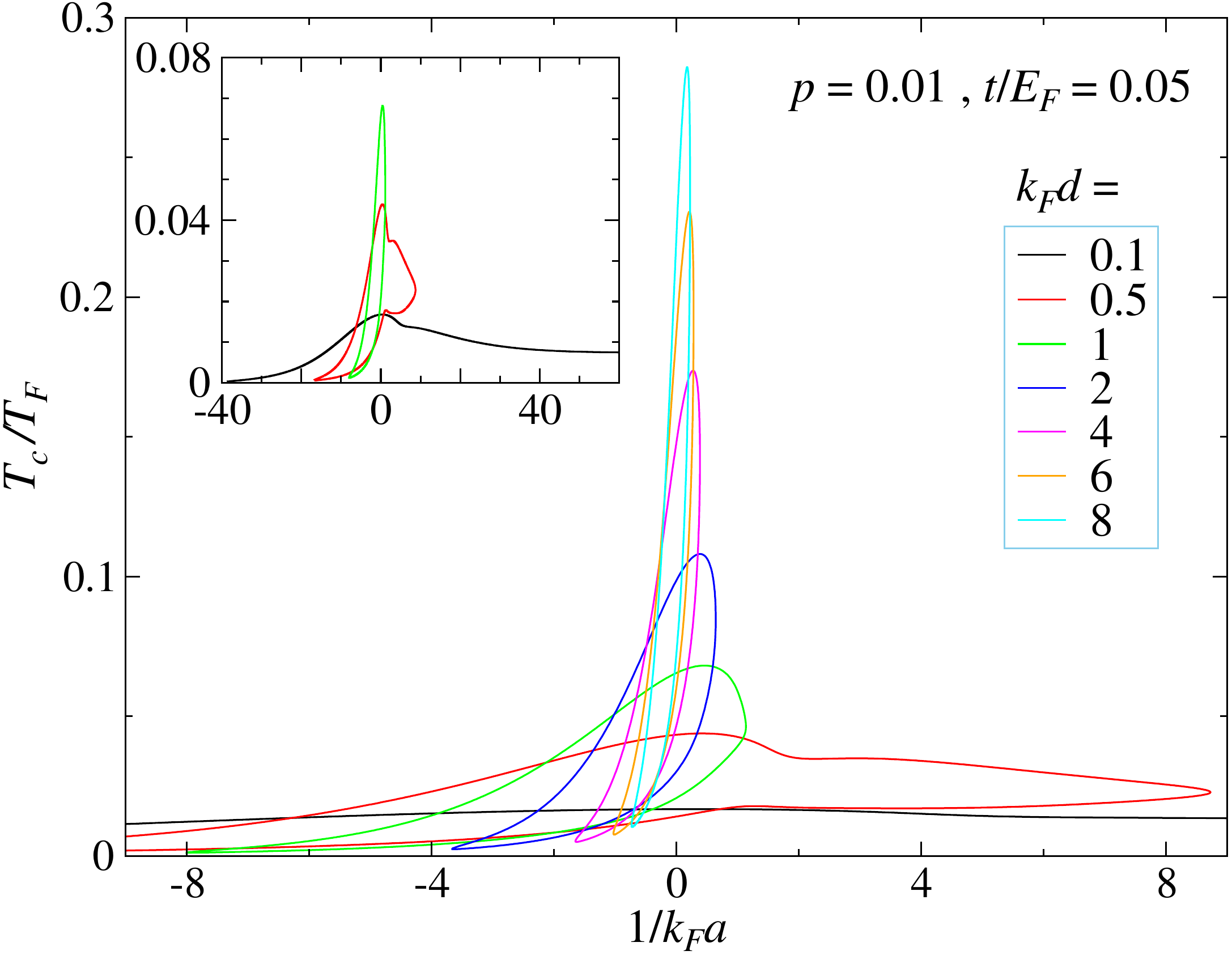}}
\caption{$T_\text{c}$ -- $1/k_\text{F}a$ phase diagram for different $k_\text{F}d$ from 0.1
  to 8 at fixed $p=0.01$ and $t/E_\text{F}=0.05$. The inset shows the small
  $d$ cases, which shares the same axis labels as the main
  figure. Increasing $d$ destroys the superfluid in the deep BEC regime.}
\label{fig:p0.01t0.05}
\end{figure}

So far, we have restricted ourselves to fairly small $d$, with
$d \leq 2$. In Fig.~\ref{fig:p0.01t0.05}, we show the behavior of
$T_\text{c}$ for a large range of $d$, from $k_\text{F}d = 0.1$ to 8, with a fixed
$p=0.01$ and $t/E_\text{F}=0.05$. For $k_\text{F}d\geq 4$, we have the range
$|k_z| <\pi/d < k_\text{F}$, which makes the lattice effect much stronger. Note
that for $t/E_\text{F}=0.05$, the $k_\text{F}d=6$ and 8 cases are physically
inaccessible. Nonetheless, these curves show a clear trend, namely,
with increasing $d$, the maximum $T_\text{c}$ increases and the $T_\text{c}$ loop
becomes narrower in terms of $1/k_\text{F}a$, more concentrated near
unitarity. On the other hand, for small $k_\text{F}d$, $\pi/d$ becomes very
large. With a small $t$ (shown in the inset), the lattice band will be
fully occupied, giving rise to an elongated open-end Fermi cylinder (for
$k_\text{F}d \ge 0.5$) in the momentum space.  Due to this small $d$, except
for the $k_\text{F}d=0.1$ case (which has a closed ellipsoid Fermi surface),
other $T_\text{c}$ curves in the figure cannot access the deep BEC
limit. Starting from a small $d$, this set of curves reveal that
increasing $d$ leads to the formation of a closed curve of $T_\text{c}$ so
that the superfluid phase in the deep BEC regime is destroyed.

From Figs.~\ref{fig:kFd1t1} to \ref{fig:p0.01t0.05}, we find that the
behavior $T_\text{c}$ has a close connection to the topology of the Fermi
surface. For a \emph{closed} Fermi surface, it can be brought into a
nearly spherical shape by momentum rescaling. For small $p$, the
situation for pairing is very much like in the 3D homogeneous
case. Therefore, the $T_\text{c}$ curve for low $p$ is similar to the 3D
homogeneous case; the lower $T_\text{c}$ vanishes in the near BEC regime, and
there exists a superfluid ground state in the BEC regime.  For
\emph{open} Fermi surfaces, pairing and superfluidity become more
difficult, making a ground state superfluid impossible. Note that for
a simple tight-binding band in the lattice dimension with
nearest-neighbor approximation, the Fermi surface topology changes
from closed below half filling to open above half filling. Above half
filling, the fermion motion on the Fermi surface becomes more
hole-like in the $k_z$ direction. While the in-plane motion is always
particle-like, this change of character may have detrimental effect on
the pairing and superfluidity.

\begin{figure}
\centerline{\includegraphics[clip,width=3.2in]{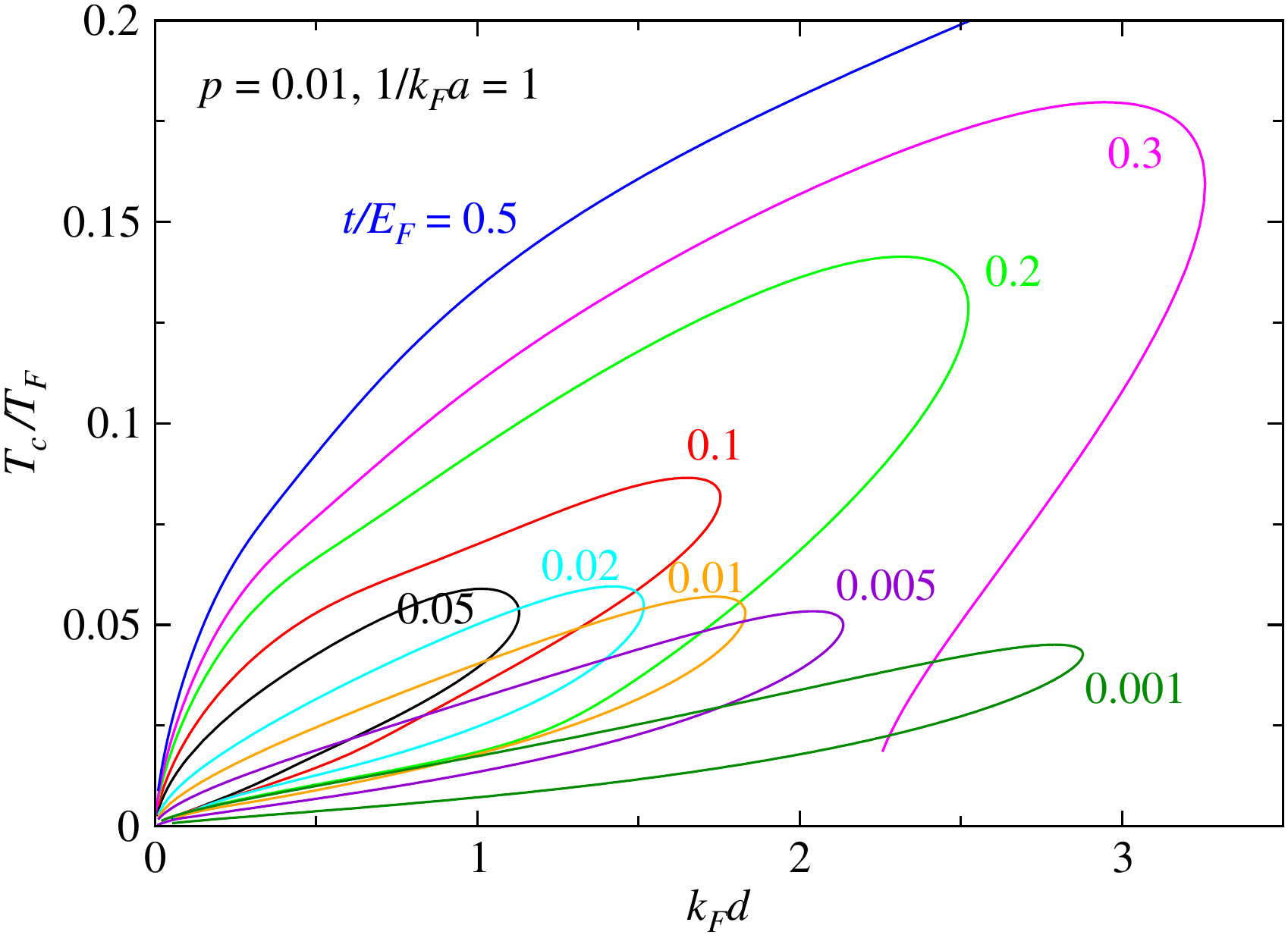}}
\caption{$T_\text{c}$ -- $d$ phase diagram for different
  $t/E_\text{F}$ from 0.001 to 0.5, as labeled, at fixed $p=0.01$ and
  $1/k_\text{F}a=1$. $T_\text{c}$ and the superfluid region both
  increase as $t$ increases.}
\label{fig:p0.01InvkFa1-d-t}
\end{figure}

\subsubsection{Continuous evolution of the superfluid phase with $t$ and $d$}

Now, we investigate how $T_\text{c}$ evolves continuously with lattice
spacing $d$. Plotted in Fig.~\ref{fig:p0.01InvkFa1-d-t} are a series
of $T_\text{c}$ curves as a function of $k_\text{F}d$, for fixed $p=0.01$ and
$1/k_\text{F}a = 1$ but different $t/E_\text{F}$ from 0.001 to 0.5. Except for the
large $t/E_\text{F}$ ($\ge 0.3$) cases , which are unphysical or hard to
realize experimentally, $T_\text{c}$ curves form a series of loops. This
agrees with the existence of two branches at this interaction
strength. The superfluid phase space area shrinks with decreasing
$t$. This means that, for small $t$ at the particular $1/k_\text{F}a = 1$, a
large $d$ will not be able to maintain the superfluid phase. At this
pairing strength, the largest reachable value of $k_\text{F}d$ is highly
nonmonotonic as a function of $t$, with a minimum of 1.13 for
$t/E_\text{F} = 0.05$. This also confirms that the ground state at
$1/k_\text{F}a = 1$ is not a superfluid for $t/E_\text{F} \le 0.2$ and $p=0.01$.
For larger $t$, the interaction parameter $1/k_\text{F}a$ at which the lower
$T_\text{c}$ would vanish becomes smaller than 1, as can be seen from
Fig.~\ref{fig:Tc-InvkFa-t}. This explains why for $t/E_\text{F} =0.3$ and 0.5
in Fig.~\ref{fig:p0.01InvkFa1-d-t}, there is no longer a lower $T_\text{c}$
solution for  $k_\text{F}d\lesssim  2.2$ and 4.1,
respectively.

\begin{figure}
\centerline{\includegraphics[clip,width=3.4in]{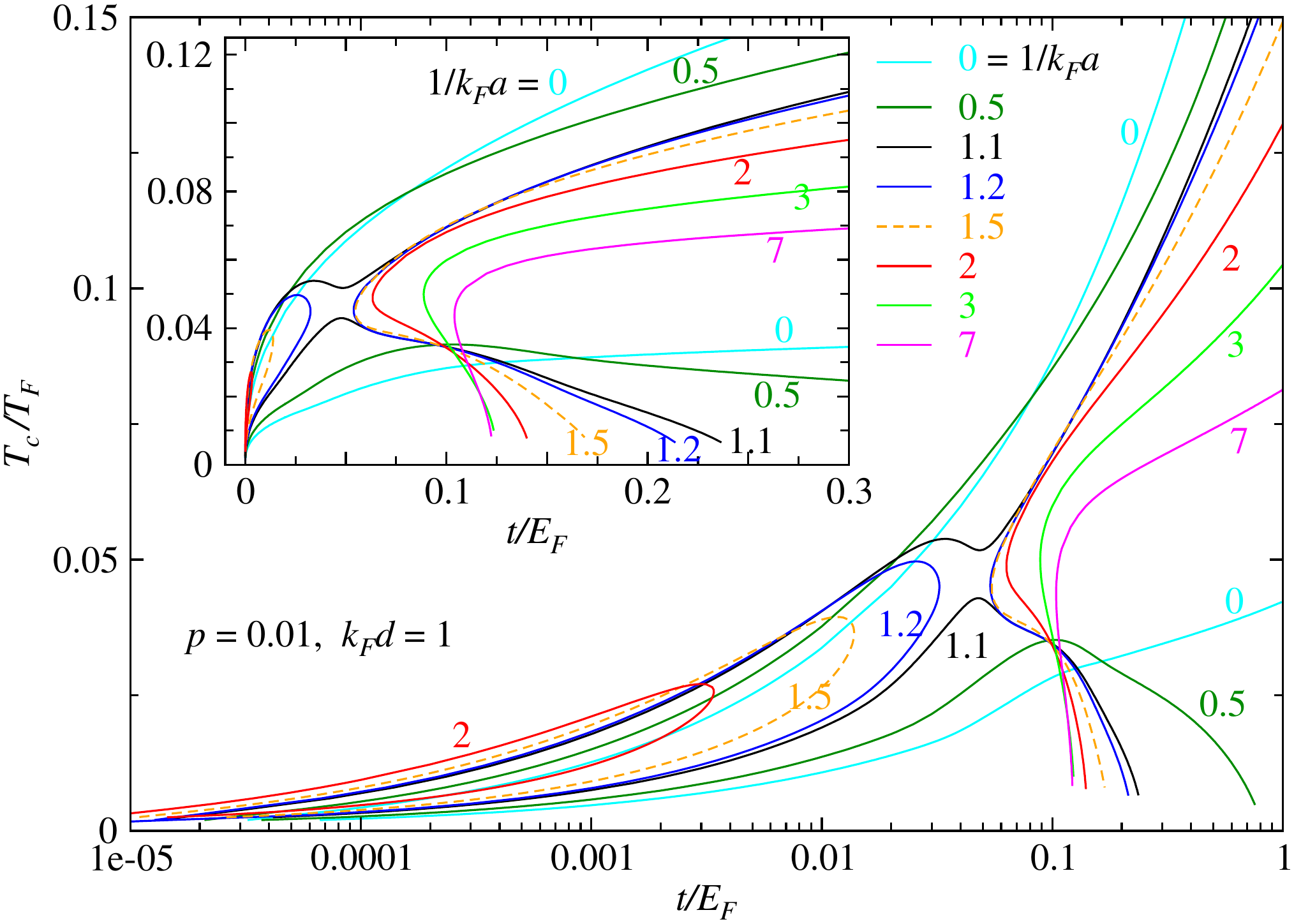}}
\caption{Behavior of $T_\text{c}$ as a function of $t/E_\text{F}$ at $p=0.01$ and
  $k_\text{F}d=1$, but for different values of $1/k_\text{F}a$, as labeled. The
  inset and the main figure share the same color coding but with
  different scales for the horizontal axis. The $T_\text{c}$ curve splits
  into two parts when $1/k_\text{F}a \ge 1.2$. }
\label{fig:Tc-t-InvkFa}
\end{figure}

The evolution of $T_\text{c}$ with continuously varying $t$ is presented in
Fig.~\ref{fig:Tc-t-InvkFa}, for a series of interaction parameter
$1/k_\text{F}a$ from 0 at unitarity to 7.0 in the BEC regime. Here $p=0.01$
and $k_\text{F}d =1$ are fixed. Logarithmic and linear scales are used for
the horizontal axis in the main figure and the inset,
respectively. The log scale serves to magnify the small $t$
regime. For $1/k_\text{F}a \le 1.1$, the curves have an upper and a lower
branch, which joins at the small $t$ end. Indeed, from Figs. 1-8, we
find that no matter whether the Fermi surface is closed or open, there
are always two $T_\text{c}$ branches in the unitary and BCS regimes.  For
$1/k_\text{F}a\ge 1.2$, we find that the $T_\text{c}$ curves pinch together and then
split into two parts around $t/E_\text{F} = 0.04$. The left part forms a
loop, which shrinks quickly as $1/k_\text{F}a$ moves towards BEC. This left
loop is the same superfluid phase as the left loop in Fig.4; they are
just different cuts of the superfluid region in the multidimensional
phase diagram. For stronger interactions in the BEC regime, either a
large $t$ or a very tiny $t$ is needed to maintain a superfluid
phase. While the former case allows a closed Fermi surface and thus a
superfluid solution in the BEC regime, the latter case will allow two
branches of $T_\text{c}$ which persist into the BEC regime. One can also tell
from this figure that, for small $t/E_\text{F}< 0.12$, either there is no
$T_\text{c}$ at all or there is a lower $T_\text{c}>0$, so that the ground state
(with $p=0.01$ and $k_\text{F}d =1$) is \emph{not} a superfluid for
$1/k_\text{F}a \le 7$.

Due to the high complexity of the multidimensional phase diagram, the
counterpart of the above figures would look somewhat different when
$(t,d,p,1/k_\text{F}a)$ changes.

\begin{figure}
\centerline{\includegraphics[clip,width=3.4in]{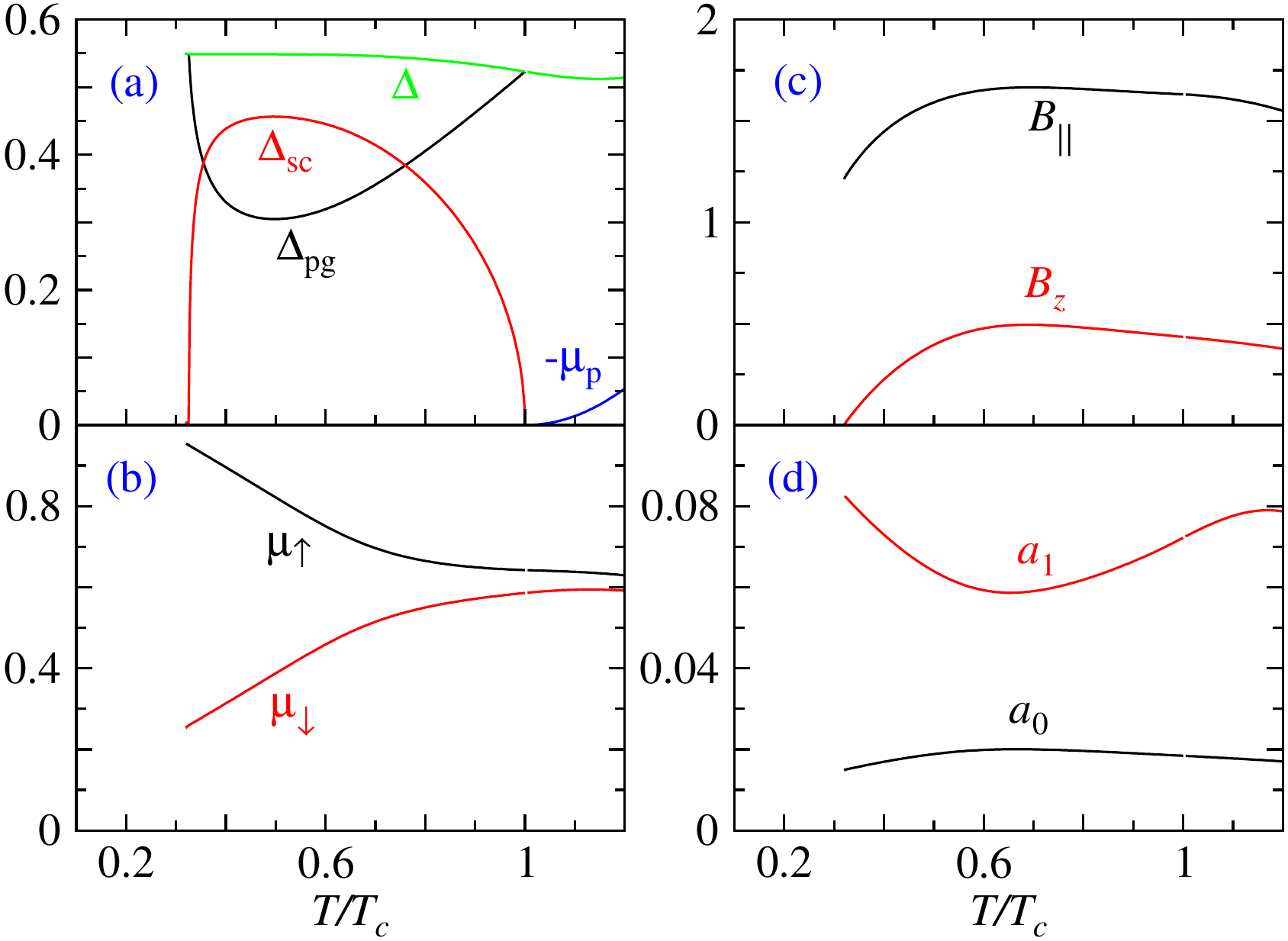}}
\caption{Behavior of (a) the gaps and $\mu_\text{p}$, (b) $\mu_\sigma$ (c)
  $B_\parallel$ and $B_z$, and (d) $a_0$ and $a_1$, as a function
  of $T/T_\text{c}$, for $t/E_\text{F} = 0.2$, $k_\text{F}d=2$ and $p=0.01$ at
  unitarity. Here $T_\text{c}/T_\text{F}=0.1736$ is the upper $T_\text{c}$, and the lower
  $T_\text{c}$ is $T_\text{c,L}/T_\text{F} = 0.05656$. Gaps and chemical potentials are
  in units of $E_\text{F}$. The coefficients $B$'s are in units of $1/2m$,
  $a_0$ and $a_1$ in units of $k_\text{F}^3/E_\text{F}^2$ and $k_\text{F}^3/E_\text{F}^3$,
  respectively. }
\label{fig:Gaps}
\end{figure}

\subsection{Gaps in the superfluid phase }
\label{subsec:Gaps}

In Fig.~\ref{fig:Gaps}, we present, as an example for intermediate
temperature superfluidity, the behavior of the order parameter
$\Delta_\text{sc}$, the pseudogap $\Delta_\text{pg}$ and the total gap $\Delta$
and a few relevant quantities as a function of temperature in the
superfluid phase. Also plotted is the solution above the upper $T_\text{c}$,
especially for the pair chemical potential $\mu_\text{p}$. Shown in the
figure is for the case of $k_\text{F}d =2$, $t/E_\text{F} = 0.2$ and $p = 0.01$ at
unitarity. It is close to the case of $t/E_\text{F} = 0.205$ in
Fig.~\ref{fig:Tc-InvkFa-t}.
Near the upper $T_\text{c}$, the behavior of the gaps look similar to regular
superfluid Fermi gases in the pseudogap regime; The order parameter
$\Delta_\text{sc}$ turns on with decreasing $T$, while the pseudogap
$\Delta_\text{pg}$ starts to decrease, leaving the total gap roughly
constant or slightly increasing. Above the upper $T_\text{c}$, the pair
chemical potential $\mu_\text{p}$ starts to decrease from 0 with increasing
$T$. The vanishing of $\Delta_\text{sc}$ at the upper $T_\text{c}$ is the same as
in BEC of ideal Bose gases. As the temperature decreases towards the
lower $T_\text{c,L}$, $\Delta_\text{pg}$ increases again, which suppresses
$\Delta_\text{sc}$ quickly down to zero. This can be understood from the
highly decreased value of $B_z = t_\text{B} d^2$ at $T_\text{c,L}$ in panel (c);
As $B_z$ decreases, pairs become heavy in the lattice direction,
leading to reduced energy cost for exciting finite momentum pairs and
hence an rapid increase in $\Delta_\text{pg}$, which then exhausts the
order parameter via $\Delta_\text{sc}^2 = \Delta^2 - \Delta_\text{pg}^2$. We
note that there are no other sharp changes in $B_\parallel$, $a_0$ and
$a_1$. Further lowering $T$ below $T_\text{c,L}$ would enter again a normal
state. However, the trend of $B_z$ at $T_\text{c,L}$ suggests that this
normal state may soon become unstable against pair density wave (or
stripe order) formation in the lattice direction (with a negative
$B_z$ at lower $T$). Other possible solutions in this low $T$ normal
state include phase separation and possible FFLO-like solutions with a
wavevector along the $\hat{z}$ direction. In fact, the pair density
wave solution is similar to an FFLO state, except that it may not
exhibit superfluidity. One would need to include the $q_z^4$ order in
the inverse $T$ matrix expansion in order to obtain a meaningful
solution below $T_\text{c,L}$, which is beyond the scope of current work.

It is interesting to note that while $\Delta$ is roughly a constant in
$T$, $\mu_\uparrow$ and $\mu_\downarrow$ becomes far apart at low
$T$. This large separation, with $h = 0.346E_\text{F}$, is comparable to the
Clogston limit for pair breaking \cite{Clogston},
$\Delta/\sqrt{2} = 0.388E_\text{F}$, where $\Delta/E_\text{F} = 0.549$ at
$T_\text{c,L}$. In other words, the disappearance of superfluidity at the
lower $T_\text{c,L}$ is compatible with the Clogston picture as well. The
small difference between $h$ and $\Delta/\sqrt{2}$ may be attributable
to the deviation of the Fermi surface from an isotropic 3D sphere
\cite{noteonClogston}. In addition, here the gap is large (beyond the
BCS regime) so that self-consistent calculations are important.  On
the other hand, at the upper $T_\text{c}$, $h$ is much smaller than
$\Delta/\sqrt{2}$, implying that the vanishing of the superfluid order
at the upper $T_\text{c}$ is not associated with the Clogston picture but
rather driven by pairing fluctuations.

\subsection{Superfluid density}

\begin{figure}
\centerline{\includegraphics[clip,width=3.4in]{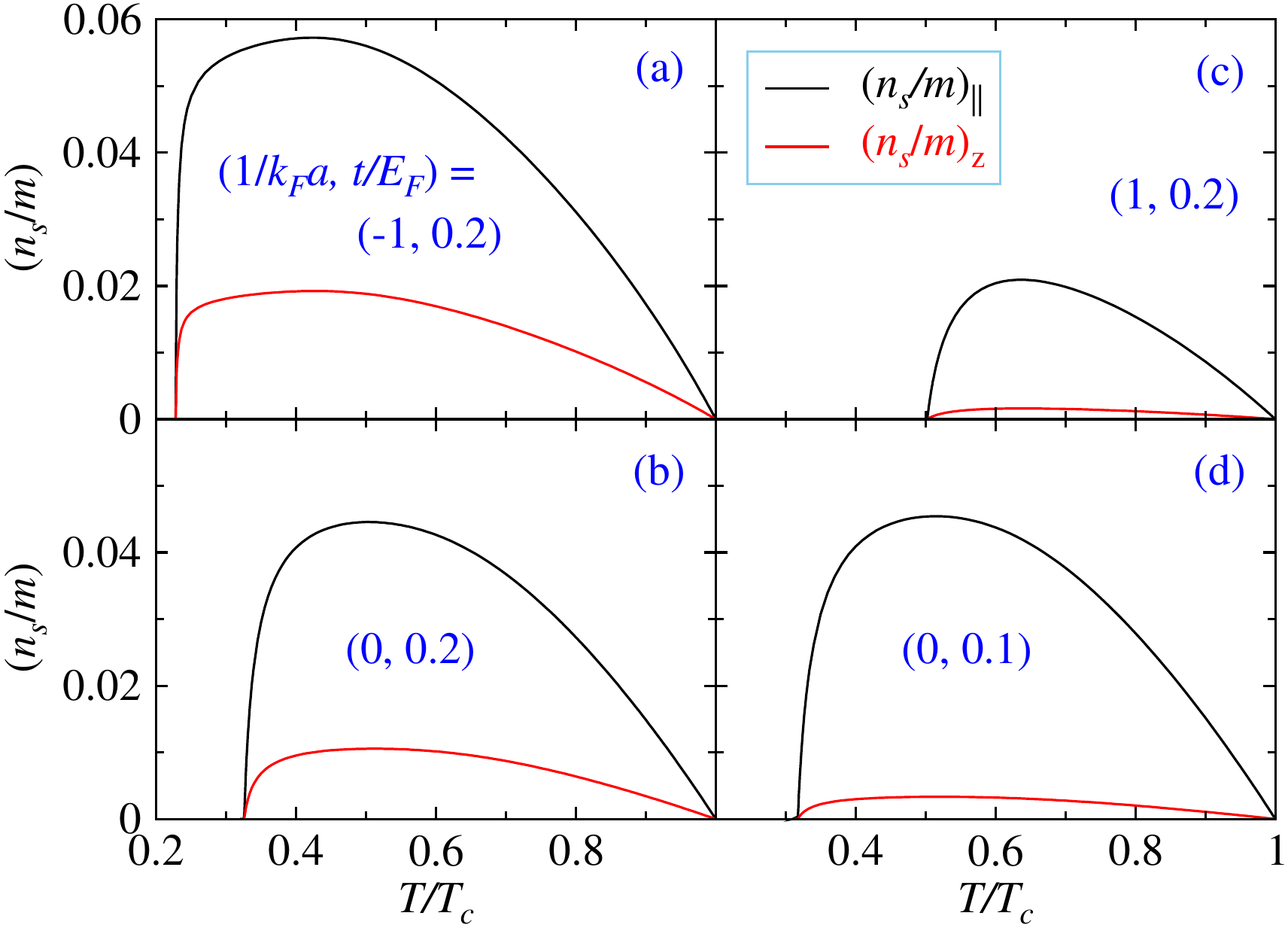}}
\caption{Behavior of the in-plane (black) and lattice components (red) of the superfluid density $(n_\text{s}/m)$ as a function of $T/T_\text{c}$, for $k_\text{F}d=2$,  $p=0.01$, and (a) $(1/k_\text{F}a, t/E_\text{F})=(-1,0.2)$, (b) (0, 0.2), (c) (1, 0.2) and (d) (0, 0.1), with $(T_\text{c}/T_\text{F}, T_\text{c,L}/T_\text{F})=(0.0923, 0.0210)$, $ (0.1736,0.0566) $, (0.1362, 0.0684), and (0.1310, 0.0416), respectively. All panels share the same legends. }
\label{fig:Ns}
\end{figure}

In this section, we show the behavior of the superfluid density. Here
we choose to show only cases of intermediate temperature
superfluidity, with both an upper $T_\text{c}$ and a lower $T_\text{c,L}$, as in Subsec. \ref{subsec:Gaps}. Cases
without a lower $T_\text{c}$ (for large $t$) are more qualitatively similar
to their balanced counterpart shown in Part I \cite{PartI}.

Plotted in Fig.~\ref{fig:Ns} are the temperature dependence of both
the in-plane (black curves) and lattice components (red curves) of
$(n_\text{s}/m)$ for $k_\text{F}d = 2$ with $p=0.01$. Panels (a-c) are for the BCS,
unitary and BEC cases, respectively, for $t/E_\text{F}=0.2$.  The
corresponding curve of $T_\text{c}$ versus $1/k_\text{F}a$ is close to the green one
for $t/E_\text{F}=0.205$ in Fig.~\ref{fig:Tc-InvkFa-t}. These results suggest
that both components decreases as $1/k_\text{F}a$ increases. The suppression
of the lattice component, $(n_\text{s}/m)_z$, can be attributed more to the
effect that the system becomes more 2D and $t_\text{B}$ decreases with
increasing pairing strength. However, the reduction of the in-plane
$(n_\text{s}/m)_\parallel$ is likely due to the increase of the pseudogap
$\Delta_\text{pg}$ with decreasing $T$ towards $T_\text{c,L}$, leading to
premature shut-off of the superfluid density before it fully reaches
its maximum possible value (normally) at $T=0$.

Shown for comparison in Fig.~\ref{fig:Ns}(d) is the case of
$t/E_\text{F}=0.1$ at unitarity, with other parameters the same as in
Fig.~\ref{fig:Ns}(b). As can be seen, the in-plane curves are very
close to each other for these two cases. However, the lattice
component is drastically suppressed by the smaller $t$ in panel
(d). This can be understood qualitatively from the increased fermion
band mass and hence the pair mass in the $\hat{z}$ direction.

For all panels in Fig.~\ref{fig:Ns}, the temperature dependencies of
both components are close to each other, despite their rather
different magnitudes. This is because the main $T$ dependence comes
from the common prefactor $\Delta^2_\text{sc}$ in Eq.~(\ref{eq:Ns}).

\subsection{BEC asymptotic behavior with $p\ne 0$}

\begin{figure}
\centerline{\includegraphics[clip,width=3.in]{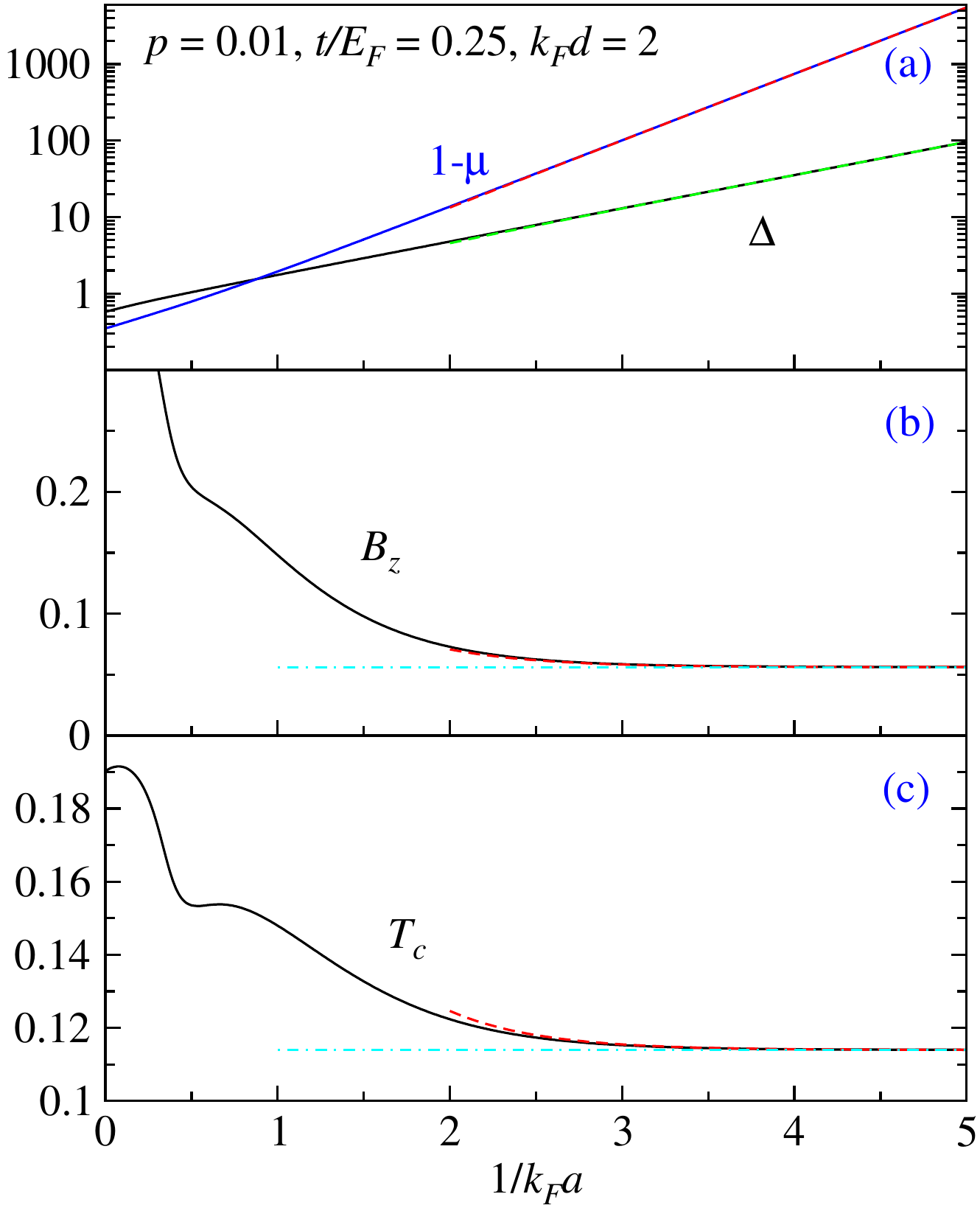}}
\caption{Behavior of (a) $1-\mu$ and $\Delta$, (b)
  $B_z$, and (c) $T_\text{c}$ as a function of $1/k_\text{F}a$, and comparison
  with the asymptotic solutions in the BEC regime. The solid lines are
  full numerical solution, and the dashed lines are the asymptotic
  solution, and the (cyan) dot-dashed lines are the BEC asymptote. Here
  $p=0.01$, $t/E_\text{F}=0.25$, and $k_\text{F}d=2$.}
\label{fig:TcBEClimit}
\end{figure}

Finally, we show in Fig.~\ref{fig:TcBEClimit} the asymptotic behavior
of $\mu$, $\Delta$, $B_z$ and $T_\text{c}$ in the BEC limit. Plotted in
Fig.~\ref{fig:TcBEClimit} are $1-\mu$ and $\Delta$ in units of $E_\text{F}$
versus $1/k_\text{F}a$ in a semi-log scale. The straight lines confirm their
exponential dependence on $1/k_\text{F}a$. The dashed lines are analytical
asymptotic solution, in perfect agreement with full numerical
solutions (solid lines). The red dashed lines in panels (b) and (c)
present the solution obtained using the asymptotic expansions, while the
cyan dot-dashed lines represent the deepest BEC asymptotes. Clearly,
the asymptotic expansions and the BEC asymptotes are all in
quantitative agreement with the full numerical solutions. This
provides direct support of our analytical derivations in the BEC
regime. These plots demonstrate that in the deep BEC limit, $B_z$
and $T_\text{c}$ approach a constant asymptote, as also shown in
Fig.~\ref{fig:t0.1d1.5}. Similar constant asymptotic behaviors are
found for $B_\parallel$, $a_0\Delta^2$ and $a_1\Delta^2$ as well, a
plot of which can be seen in Ref.~\cite{1DOLshort}.

\begin{figure}
\centerline{\includegraphics[clip,width=3.4in]{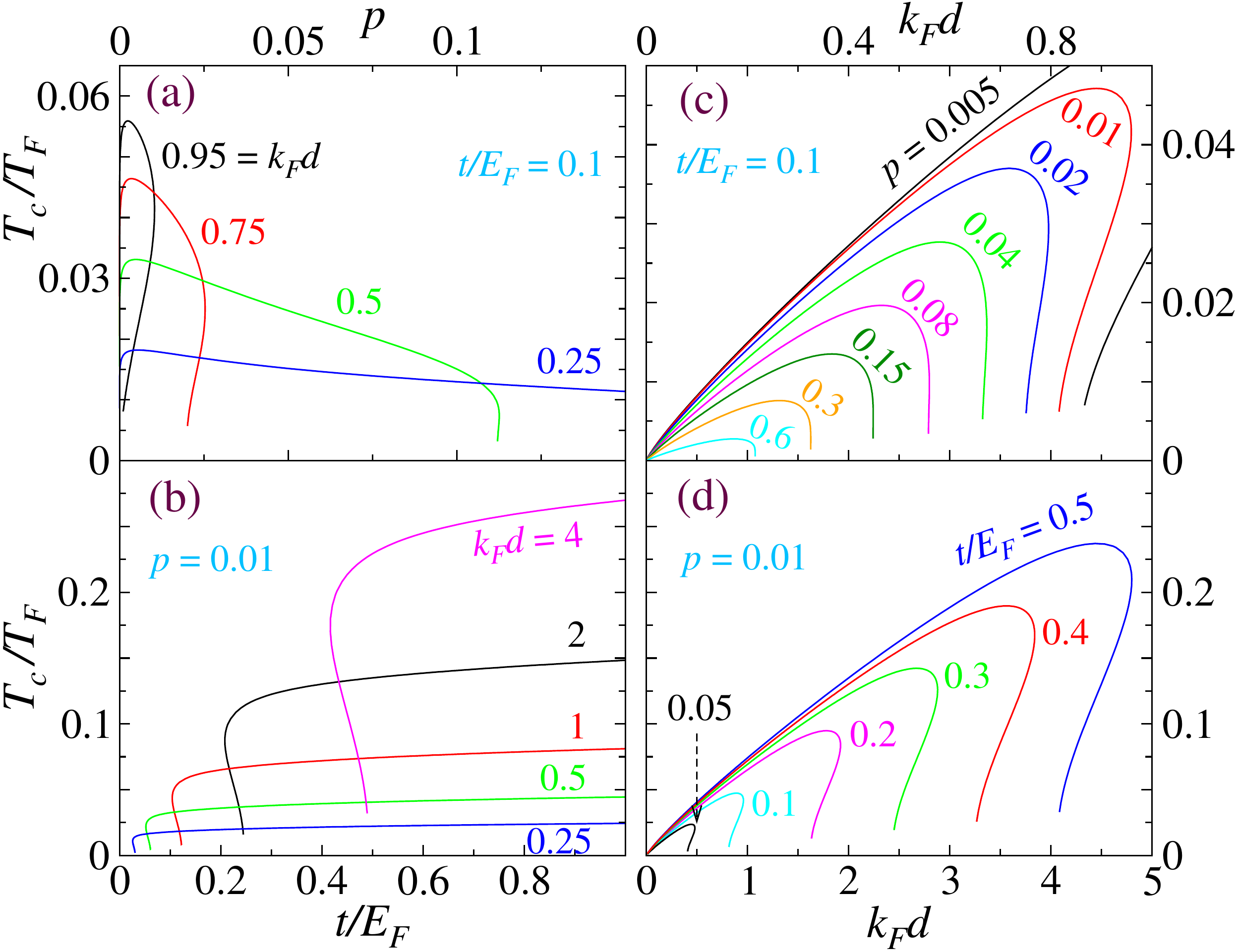}}
\caption{Behavior of the BEC asymptote of $T_\text{c}$, as a function (a) of $p$
  for fixed $t/E_\text{F} = 0.1$ with different $k_\text{F}d$ from 0.25 to 0.95, (b)
  of $t/E_\text{F}$ for fixed $p=0.01$ with varying $k_\text{F}d$ from 0.25 to 4,
  (c,d) of $k_\text{F}d$ for (c) fixed $t/E_\text{F} = 0.1$ with varying $p$ from
  0.005 to 0.6 and for (d) fixed $p=0.01$ with varying $t/E_\text{F}$ from 0.05
  to 0.5. The parameters are labeled on the curves. }
\label{fig:TcBECAsymptote}
\end{figure}

Given these BEC asymptotic behaviors, we can investigate the phase
diagrams in the BEC limit as a function of $t$, $d$ and $p$.  Shown in
Fig.~\ref{fig:TcBECAsymptote}(a) is the BEC asymptote of $T_\text{c}$ with
$t/E_\text{F}=0.1$ as a function of $p$ with different $k_\text{F}d = 0.25$, 0.5,
0.75 and 0.95.  The Fermi surface topology at $p=0$ changes from closed to
open, as the lattice spacing increases across $k_\text{F}d=
0.942$. Therefore, nearly all cases shown here have a closed
Fermi surface.  It can be readily seen that for $k_\text{F}d = 0.95$, the
maximum $p$ is only about 0.01; there is no BEC superfluid solution
for larger $p$. The maximum $p$ increases as $d$ decreases. For
smaller $k_\text{F}d = 0.25$, $p$ survives up to about 0.475. This may
largely have to do with the fact that a smaller $d$ places less
restrictive confinement for pair motion in the $k_z$ direction, and
thus the system is closer to the 3D case, so that it can accommodate a
larger population imbalance.

Plotted in Fig.~\ref{fig:TcBECAsymptote}(b) are $T_\text{c}$ curves with
$p=0.01$ as a function of $t/E_\text{F}$ with different $k_\text{F}d = 4$, 2, 1, 0.5
and down to 0.25. These curves demonstrate that the lowest threshold
of $t$ for having a BEC superfluid solution increases with $d$. For
$k_\text{F}d=4$, we need $t/E_\text{F} \gtrsim 0.42$. For $k_\text{F}d = 2$, the threshold
drops to about 0.21, in agreement with Fig.~\ref{fig:Tc-InvkFa-t}. For
$k_\text{F}d=0.25$, the threshold becomes $t/E_\text{F}\approx 0.03$. In particular,
for $k_\text{F}d = 1$, the threshold is about 0.105 ($>0.1$). This explains
why there is no $k_\text{F}d=1$ curve in panel (a), calculated for
$t/E_\text{F}=0.1$.  As $d$ increases, the overall $T_\text{c}$ also increases,
since the 2D planar density $n_\text{2D}$ increases and so does the
noninteracting chemical potential.  In reality, $t$ is normally
small. This requires a small $d$ in order to have a BEC superfluid, as
one can see from Fig.~\ref{fig:p0.01t0.05} as an example, where only
the $k_\text{F}d=0.1$ curve persists into the BEC limit for small
$t/E_\text{F}=0.05$. Our calculations show that these thresholds roughly
correspond to half filling of the lattice band, where the Fermi surface
topology changes.

Presented in Fig.~\ref{fig:TcBECAsymptote}(c) is the BEC asymptote of
$T_\text{c}$ calculated for $t/E_\text{F} = 0.1$, as a function of $k_\text{F}d$ with
different population imbalances from $p = 0.005$ to 0.6. The maximum
allowed $k_\text{F}d$ decreases quickly with increasing $p$. For $p = 0.005$,
$k_\text{F}d$ goes up to 1.2. For $p=0.01$, $k_\text{F}d$ is allowed up to about
0.96. For $p=0.6$, one needs a small $k_\text{F}d < 0.22$ to have a BEC
superfluid.  Figure \ref{fig:TcBECAsymptote}(c) also reveals
that for a given $d$, there is a maximum allowed $p$, beyond which the
BEC superfluid solution no longer exists, in agreement with
Fig.~\ref{fig:TcBECAsymptote}(a).

Shown in Fig.~\ref{fig:TcBECAsymptote}(d) is the BEC asymptote of
$T_\text{c}$ calculated for $p = 0.01$, as a function of $k_\text{F}d$ with
different tunneling $t/E_\text{F}$ from 0.05 to 0.5. As is shown, the maximum
possible $d$ increases with $t$. While for $t/E_\text{F} = 0.5$ this maximum
is about 4.8, it decreases down to about 0.49 for $t/E_\text{F} = 0.05$. If
one wants to have a larger $d$ for the same small $t$, one will need
to use a smaller $p$, as indicated by
Fig.~\ref{fig:TcBECAsymptote}(c). The lower end $t/E_\text{F} = 0.05$ is more
realistic. It means that for a typical $k_\text{F}d \sim 1$, a small amount
of population imbalance will be sufficient to destroy the superfluid
solutions in the BEC regime \cite{noteonTc}.

We point out that in all four panels of Fig.~\ref{fig:TcBECAsymptote},
there exists a narrow range of the parameters where the $T_\text{c}$
curve bends back and thus is double-valued, which correspond to the
two branches such as the low $p$ curves shown in
Fig.~\ref{fig:t0.1d1.5}(e), with an open Fermi surface. For the rest
part of the curves, there is only one (upper) $T_\text{c}$,
corresponding to, e.g., the low $p$ curves in
Fig.~\ref{fig:kFd0.5t0.1}, with a closed Fermi surface.


\subsection{Further Discussions}

From the numerical results presented above, we see that the behavior
of $T_\text{c}$ and the phase diagrams are very complex, in the presence of a
population imbalance. In the physically accessible scope of the
parameters, e.g., constrained by the condition $2m td^2 <1$, the
superfluid phase occupies only a very restricted small volume in the
multi-dimensional phase space. Superfluidity can be easily destroyed
by small amount of population imbalance when the lattice constant $d$
becomes large and/or the tunneling matrix element $t$ becomes
small. To understand this destruction of superfluidity, we notice that
large $d$ and small $t$ put the system in the quasi-2D regime, such
that the lattice band is essentially fully occupied, and in-plane
chemical potential (in the noninteracting limit) is much higher than
the lattice band width $2t$, leaving almost no dispersion on the Fermi
surface along the lattice direction. Excessive fermions will
necessarily have to occupy high in-plane momentum states and thus cost
a lot of excitation energy. In this case, a small population imbalance
will create a substantial mismatch $h$ in chemical potentials that is
sufficient to destroy pairing.

On the other hand, we find that smaller $d$ is more benign in the
behavior of $T_\text{c}$, e.g., the $k_\text{F}d=0.1$ case in
Fig.~\ref{fig:p0.01t0.05}. For small $d$, the momentum space
constraint $|k_z|<\pi/d$ in the lattice direction becomes less
restrictive so that the Fermi surface becomes an ellipsoid, which can
be mapped back into a sphere via momentum rescaling. Whether closed or
open, the Fermi surface topology in the noninteracting limit plays an
important role in classifying the behavior of the $T_\text{c}$ curves. With a
closed Fermi surface, the superfluid solution in the BEC regime (if it
exists) has only one (upper) $T_\text{c}$. In contrast, with an open Fermi
surface, the superfluid has both an upper and a lower $T_\text{c}$. Further
careful analysis may involve different Fermi surfaces for the two spin
components and how their influence evolves with $p$.

More surprisingly, when the superfluid solution exists in the BEC
regime or on the BEC side of unitarity, $T_\text{c}$ can be substantially
enhanced by a small amount of population imbalance with respect to the
balanced case. Via analytical analysis in the BEC regime, we show that
this enhancement is associated with contributions to $t_\text{B}$ from
excessive unpaired majority atoms. These contributions lead to a
constant BEC asymptote for $t_\text{B}$ and a few other quantities, and hence
a constant BEC asymptote for $T_\text{c}$ via the pseudogap equation. These
contributions to $t_\text{B}$ constitute a new pair hopping mechanism
assisted by excessive majority atoms. For this mechanism to work, it
is important that there is at least one transverse continuum
dimension. In the present case of 1D optical lattice, there are two
transverse continuum dimensions, i.e., the 2D $xy$ plane. This
guarantees that there are always excessive majority atoms available on
a neighboring lattice ``site''. Therefore, lattice-continuum mixing is
crucial for this unusual behavior.

Another important difference between 1D optical lattices and the 3D
continuum case is the pair fraction in the BEC limit. For the latter
case, all minority atoms will form pairs, namely,
$n_\text{p}/n_\downarrow = 1$ in the BEC limit. In contrast, for the present
case, we always have $n_\text{p}/n_\downarrow < 1$ for nonzero $p$, as can be
seen from Eq.~(\ref{eq:a0BECp}). The difference can be attributed to
the quasi-two dimensionality in the present case, which leads to a
constant ratio of $\Delta^2/|\mu|$ in the BEC limit, in contrast to
vanishing as $1/\sqrt{|\mu|}$ in 3D continuum.

Our calculations are based on the assumption that the 2D planes are
homogeneous. In real experiments, they are always finite and confined
in a shallow trapping potential. At the same time, the lattice
direction is confined by a trapping potential as well.  The finite
size and trap effects are beyond the scope of the current work and
will be left for future investigations. We note that recent progress
in implementing uniform box trapping potential \cite{PhysRevLett.110.200406,PhysRevLett.118.123401,PhysRevLett.120.060402} can greatly reduce the
complexity. 

\section{Conclusions}

In summary, we have studied the ultracold atomic Fermi gases in a 1D
optical lattice in the presence of population imbalance with a pairing
fluctuation theory, as they undergo a BCS-BEC crossover. We find that
superfluidity exists only for a very restricted range of parameters,
while it can be readily destroyed by a small amount of imbalance $p$
at large $d$ and small $t$.  When the superfluid solution does exist
on the BEC side of the Feshbach resonance, $T_\text{c}$ can be enhanced
substantially by even a tiny amount of population imbalance, via the
new pair hopping mechanism assisted by excessive majority atoms.  In
general, when $t$ is small, the $T_\text{c}$ curve bends back on the BEC side
and the superfluidity disappears in the deep BEC regime.  Meanwhile,
the superfluid phase shrinks as $p$ increases.  For fixed $d$ and $p$,
the superfluid region in the $T$ -- $1/k_\text{F}a$ plane shrinks as $t$
decreases, while for fixed $p$ and $t$, the $T_\text{c}$ curve forms a closed
loop in the $T$ -- $1/k_\text{F}a$ plane and becomes narrower near unitarity
as $d$ increases.  In general, whether there is only one (upper) $T_\text{c}$
or there are both an upper and a lower $T_\text{c}$ in the BEC regime depends
largely on the Fermi surface topology. The former occurs with a closed
ellipsoidal Fermi surface, while the latter happens when the Fermi
surface has two open ends at the Brillouin zone boundaries. Further
more, due to the quasi-two dimensionality, only part of the minority
atoms will be paired even if superfluidity exists in the BEC limit.

Our results demonstrate that experimentally one needs to be careful to
maintain a good population balance to stay in the superfluid phase. On
the other hand, a perfect balance may not always be desirable. A small
amount of imbalance may be good for enhancing $T_\text{c}$, making the
superfluid phase easier to access. It may take some trial and error
to find the optimal parameters in experiment.

These predicted behaviors of fermions on a 1D optical lattice are very
different from pure 3D continuum or 3D lattices, and have not been
reported in the literature. Since optical lattices have been realized
experimentally for a long time, these predictions should be tested in
future experiments.


\section{Acknowledgments}
We thank the useful discussions with  Chenchao Xu and Yanming
Che. This work was supported by the NSF of China (Grant No. 11774309
and No. 11674283), and the NSF of Zhejiang Province of China (Grant
No. LZ13A040001).  C. Lee was supported by the Key-Area Research and
Development Program of GuangDong Province under Grants
No. 2019B030330001, the NSF of China under Grants No. 11874434 and
No. 11574405, and the Science and Technology Program of Guangzhou
(China) under Grant No. 201904020024.


\begin{thebibliography}{37}
\bibitem [{not({\natexlab{a}})}]{notesonFFLO}%
  \BibitemOpen
  \href@noop {} {} \bibinfo {note} {Solution of the
  Fulde-Ferrell-Larkin-Ovchinnikov (FFLO) state \cite{FF,LO} has also been
  contemplated in the ``Unstable" region of the phase diagram. However, the
  FFLO states are known to be mostly unstable \cite{WangLOFF}.}\BibitemShut
  {Stop}%
\bibitem [{not({\natexlab{b}})}]{noteonClogston}%
  \BibitemOpen
  \href@noop {} {}  \bibinfo {note} {In principle, here the
  gap $\Delta$ should be the one calculated in the absence of imbalance in the
  original Clogston derivation. For $p = 0.01$, the gap is close to its $p=0$
  value.}\BibitemShut {Stop}%
\bibitem [{not({\natexlab{c}})}]{noteonTc}%
  \BibitemOpen
  \href@noop {} {}  \bibinfo {note} {We note that in
  Fig.~\ref{fig:TcBECAsymptote} most $T_c$ curves do not extend down to zero
  $T$. This happens mainly because $B_z$ becomes exponentially small, causing
  round-off errors and making it difficult to obtain a precise solution
  numerically.}\BibitemShut {Stop}%
\end{thebibliography}

%

\end{document}